\documentclass{article}

\usepackage[english]{babel}
\usepackage[a4paper,top=2cm,bottom=2cm,left=3cm,right=3cm,marginparwidth=1.75cm]{geometry}

\usepackage{amsmath}
\usepackage[makeroom]{cancel}
\usepackage{bbm}
\usepackage{mathtools}
\usepackage{empheq}
\usepackage{amssymb}
\usepackage{array}
\usepackage{bbold}
\usepackage{caption}
\usepackage{graphicx}
\usepackage{subcaption}
\usepackage{multirow}
\usepackage[utf8]{inputenc}
\usepackage[T1]{fontenc}
\usepackage{lmodern}
\usepackage[rgb]{xcolor}
\usepackage[author={Come ANNICCHIARICO}]{pdfcomment}
\allowdisplaybreaks
\usepackage{afterpage}
\usepackage{wrapfig}
\usepackage{epstopdf}
\usepackage{svg}
\usepackage{authblk}

\DeclareMathOperator*{\argmin}{\arg\!\min}

\begin{document}
\title{An Active Inference perspective on Neurofeedback Training}

\author[1,2]{Co\^me ANNICCHIARICO}
\author[2]{Fabien LOTTE}
\author[1]{J\'er\'emie MATTOUT}
\affil[1]{COPHY Team, Lyon Neuroscience Research Center, CRNL, INSERM, U1028, Lyon, FRANCE}
\affil[2]{POTIOC Team, INRIA centre at the University of Bordeaux / LaBRI, Talence, FRANCE}

\maketitle

\begin{abstract}
Neurofeedback training (NFT) aims to teach self-regulation of brain activity through real-time feedback, but suffers from highly variable outcomes and poorly understood mechanisms, hampering its validation. To address these issues, we propose a formal computational model of the NFT closed loop. Using Active Inference, a Bayesian framework modelling perception, action, and learning, we simulate agents interacting with an NFT environment. This enables us to test the impact of design choices (e.g., feedback quality, biomarker validity) and subject factors (e.g., prior beliefs) on training. Simulations show that training effectiveness is sensitive to feedback noise or bias, and to prior beliefs (highlighting the importance of guiding instructions), but also reveal that perfect feedback is insufficient to guarantee high performance. This approach provides a tool for assessing and predicting NFT variability, interpret empirical data, and potentially develop personalized training protocols.

\end{abstract}

\section{Introduction}\label{sec:intro}

Brain-Computer Interfaces (BCIs) encompass a large number of technologies specialized in acquiring brain signals, analyzing them and translating them into commands or information for an external system \cite{Shih2012}. Neurofeedback is a specific use-case of BCIs in which the signals, typically acquired using electroencephalography (EEG), are aimed at promoting the self-modulation of brain activity. The ability of animals to control their own brain activity through Neurofeedback-driven operant-conditioning has been established as early as the 1970's \cite{fetz_operant_1969}. Beyond those first experiments, the nature of the learning mechanisms that would allow a human to control a BCI is still very much debated \cite{Wood2014,Hiremath2015,sitaram_closed-loop_2017,Corsi2020}. In fact, effective control of such interfaces has proven very difficult or even impossible for a significant proportion of users leading to the emergence of the debated concepts of BCI illiteracy \cite{Vidaurre2010,Thompson2019} and Neurofeedback non-responders \cite{Alkoby2018}.\\

\begin{figure}
    \includegraphics[width=\linewidth]{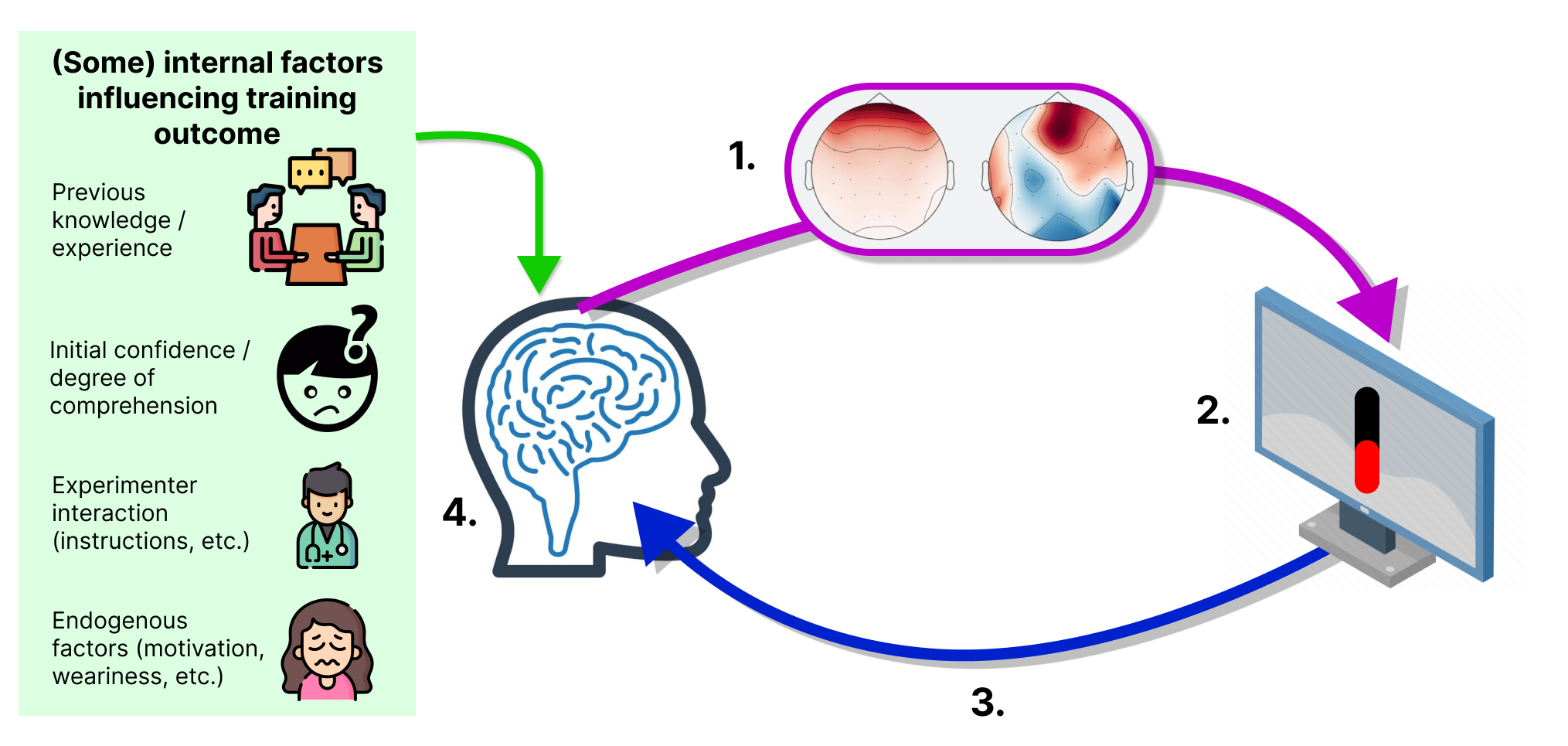}
    \caption{A schematic representation of the neurofeedback closed-loop paradigm. \textbf{1.} The subject physiological brain signals (electromagnetic, BOLD signal, etc.) are acquired using EEG, MEG, fMRI, etc.\textbf{2.} They are used to infer the hidden subject cognitive state that caused it. The experimenter derives a task-specific feedback from the infered state. Note that this latent state inference relies on strong hypotheses made by the experimenter about parts of the neurophysiological / measuring process underlying the neurofeedback. \textbf{3.} The subject perceives the feedback through various sensory means. (auditory, visual, etc.) and tries to relate it to its own cognitive state (\textbf{4.}).The subject thus tries to learn the relationship between his hidden cognitive states and the indicator displayed on the screen. On the left, a few other factors which may influence the cognitive dynamics of the subject training are shown.} 
    \label{mazex_explo_fig}
\end{figure}

Neurofeedback training attempts to teach subjects how to control specific biomarkers derived from their own brain activity (or neuromarker). A neuromarker designates a measurable neurophysiological activity that is assumed to specifically and truthfully reflect a psychological or mental state of interest.  The most widespread Neurofeedback protocols featuring EEG measurements target spontaneous brain rhythms, that is oscillations in specific frequency bands \cite{marzbani_methodological_2016,arns_neurofeedback_2017}, with a relatively loose spatial resolution. Other, less common approaches involve different brain markers (such as connectivity estimates) or more involved measuring apparati like functional Magnetic Resonance Imaging (fMRI) \cite{bagherzadeh_alpha_2020} or magnetoencephalography (MEG) \cite{Okazaki2015-oe}, with better spatial resolution but narrower out-of-the-lab application perspectives. \\

\cite{enriquez-geppert_neurofeedback_2019} enumerates three major domains of application of Neurofeedback. First, NFT training has been used by researchers to improve healthy subjects cognitive performances \cite{GRUZELIER2014124,gruzelier_2014_b,yamashita_2017}. This approach relies on brain plasticity to initiate directional changes in healthy subjects behaviour, in areas such as attention, spelling, confidence, etc. \cite{Loriette2021}. Another emerging field of use of NFT is as a scientific tool \cite{sitaram_closed-loop_2017}. Indeed, Neurofeedback can be used to favour the emergence of specific neurophysiological patterns and study the associated behavioural ones \cite{bagherzadeh_alpha_2020} as well as the associated subjective reports. Most neurofeedback studies however focus on the development of non-pharmacological therapies for psychiatric and neurological disorders such as Attentional Deficit and Hyperactivity Disorder (ADHD).  \cite{enriquez-geppert_neurofeedback_2019,Arns2020-mb,Hasslinger2020,vandoren_2019}, insomnia \cite{Hauri1982,schabus_better_2017,lambert_2021,hammer_2011,Li2022}, anxiety \cite{Moradi2011,Hou2021,Chen2021}, epilepsy \cite{Zhao2009,Strehl2014}, chronic pain \cite{Roy2020}, etc. Such NFT paradigms usually aim at having a significant (positive) impact on behavioral symptoms (e.g. focusing and learning abilities in children with attention disorder). \\

Wide-scale adoption of neurofeedback protocols for therapy by the general public has radically increased in the last decade, yet no consensus on the efficacy of neurofeedback therapies has been reached by the scientific community \cite{lubar_evaluation_1995,micoulaud-franchi_eeg_2014,cortese_neurofeedback_2016}. One of the principal challenges in evaluating the overall efficacy of NFT lies in its broad therapeutic claims, which have led to its application across a diverse range of biomarkers. This diversity complicates the comparison of findings across different studies. Furthermore, the disorder-specific relevance of these biomarkers is often inadequately validated \cite{batail_eeg_2019}. Typically, researchers rely on empirical evidence showing a significant correlation between a cognitive disorder (e.g., attention deficit) and a specific neuromarker deviation (e.g., the theta/beta power ratio as measured by EEG). They hypothesize that normalizing this deviation will consequently enhance cognitive function in the affected individuals. To test these hypotheses and disambiguate effects beyond placebo or non-specific training factors, double-blind randomized control trials have become the gold standard in Neurofeedback studies. However, randomized controlled trials are particularly cumbersome to set up. The outcome of existing studies are difficult to synthesize as they differ in many aspects including the targeted population and biomarkers, the design and protocol used, etc. Moreover, sample size is often limited. As a result, evidence is scarce and fragile \cite{marzbani_methodological_2016,enriquez-geppert_neurofeedback_2019,Trambaiolli_study_2020,riesco_2021}. For example, although neurofeedback therapy for ADHD patients has become common practice, several studies concluded that changes in subjects' behaviour cannot be proved to be more than a placebo effect \cite{Leins2007,Arns2013,Chronis2013}. Even if the training is successful on the short term, the long-term effects of neurofeedback therapies are often lackluster and in need of complementary solutions \cite{molina_mta_2009}. This has led to debates regarding how experiments should be performed: what is the aim of the training, how to implement a proper control condition e.g. with sham feedback, how to evaluate the effect and specificity of the training, etc. \cite{vandong_2013,Cannon2014ThePO,vandongen_2014}. \\

In recent years, although a few experimental studies and meta-analysis concluded in favor of encouraging results \cite{Arns2009,arns_neurofeedback_2017,Pimenta2021}, others pointed the empty part of the glass.  This obvious difficulty faced by the community to draw solid and common conclusions, has led to a call for protocol and reporting standardization \cite{micoulaud-franchi_eeg_2014,Thibault2018,ros_consensus_2020}. Moreover, to explain the inconsistency of the results, researchers have also shown interest in comparing methodological and technical details of those approaches and proposing different explanations for success or failure \cite{marzbani_methodological_2016,cortese_neurofeedback_2016,Bussalb2019}.\\

This joint effort and strong call for strengthening our methodology is very encouraging and a clear sign of maturation of the young science of neurofeedback. However, it still lacks a common generic description of the training process, used to clearly articulate our hypotheses around our partial understanding of the mechanism at play during NFT. This missing piece is a mandatory next step to further guide the development of neurofeedback, as it should act as a common language for researchers to \textbf{1. understand} what are the fundamental goals of NFT, \textbf{2. formalize} what hypotheses are made when designing a training paradigm, \textbf{3. quantify} the effect of explicit and implicit training factors in order to predict the outcome and \textbf{4. communicate} their results and interpret the ability (or inability) of the subjects to control their brain activity. This paper shows that the Active Inference framework\cite{friston_active_2016}, briefly outlined in \ref{section:ai_theory}, can be leveraged to provide a complete description of BCI / NF training. In the remainder of this introduction, we give the reader an overview of the main components of a neurofeedback training loop and of the main factors explaining the variability between outcomes. Then, we give a quick overview of existing modeling works that tackled the NFT problem. Finally, we motivate the introduction of a novel high-level subject-centered model of BCI interaction to account for all the previously explained factors.\\

\textbf{Neurofeedback uncertainties and hypotheses : }Neurofeedback training relies on numerous functional components. There have been efforts by the scientific community to formalize which elements of the neurofeedback pipeline could explain training failures. \cite{arns_neurofeedback_2017} summarizes the main advantages and pitfalls of neurofeedback in children with ADHD. The authors point out two categories of issues that mostly affect the efficacy of the training: technical ones related to recordings or trial design parameters (signal quality, processing algorithms, reward threshold and timing, feedback nature, session structure and frequency, trial-based and session-based learning curves, experimenter's influence, transfer exercises...) and issues related to learning mechanisms (learnability, perceptibility, mastery, motivation, autonomy). Predictive approaches such as those introduced in \cite{Alkoby2018,Weber2020} focus on figuring out which neurophysiological markers could explain training success or failure, and which mechanisms explain the differences of outcomes between subjects. \\

Finally, \cite{batail_eeg_2019} propose several intrinsic factors challenging neurofeedback efficacy and slowing its recognition as a valid therapeutic method by the larger neuroscientific community. First is the biomarker hypothesis used to infer subject cognitive states. We still struggle to understand the exact relationship between brain activity features and cognitive dimensions. It is also challenging to understand how a mental disorder affects brain activity. This limits attempting to use brain-wide physiological signals as a vector for therapy requires to specific, reliable and observable marker \cite{Yamada2017}. Second, a lot of neurofeedback training factors are subject-specific. We have a very superficial knowledge on how to best fit the neurofeedback loop to the subject, and the effect of endogenous factors (motivation, attention, drowsiness,etc.) and exogenous factors (effect of instructions, task set, feedback design) is still hardly quantified. Finally, there is a significant lack of understanding of the mechanisms behind conscious regulation of brain mechanisms. This broad expression hides a semantic ambiguity between two key questions with regard to the practice of neurofeedback : (1.) \textit{How} is the subject learning ? and (2.) \textit{What} is the subject learning ? Although related, they tackle different angles of the paradigm : 
\begin{itemize}
    \item "How is the subject learning ?" targets the pseudo-algorithm that best describes subject knowledge gathering during a training session. The nature of most NF exercises as exploratory tasks with the goal of optimizing a positive feedback naturally points towards operant conditioning (Skinner) \cite{Gevensleben2009a}. Subjects favor actions that led to positive (changes in the) feedback, and refrain from actions that produced opposite effects. However, operant conditioning alone struggle to tackle the issue of generalization after training, in the absence of an explicit reward. An alternative view, that we will showcase in this paper, formalizes the training as learning an efficient representation of the full neurofeedback loop. Instead of learning what actions lead to positive feedbacks, subjects learn how their actions affected their (mental) environment, with increasing the feedback being just part of their drive.
    \item "What is the subject learning ?" focuses on the goal of the training and the effect of the sessions on the dynamics of the subject. Although intimately linked to the previous point, it focuses on what part of the subject dynamics are affected by the training. In operant conditioning,  what is learnt is often the mapping between a stimulus (e.g. a low level of feedback) and an action probability. This is somewhat limited as it struggles to explain why such a training would have any usefulness in the absence of an explicit feedback (in effect reducing the true NF effect to a simple habit). Two potentially compatible elements may mitigate this fact : first, NFT may be a way for the subject to learn how to interpret implicit (interoceptive) feedback signals (heart rate, metacognitive awareness, etc.). By associating these signals with the performed actions during training (following some kind of classical conditionning (Pavlov)), the subject becomes able to use this knowledge outside of the training environment. Second, subjects may learn how actions affect hidden states rather than what actions to perform given a specific feedback. This is in line with \cite{veilahti_neurofeedback_2021} who argue that skill learning \cite{Gevensleben2009}, defined as a conscious and active mechanism of model building following observations, plays an integral part during neurofeedback training.
\end{itemize}

\textbf{Factors of neurofeedback efficacy :} Numerous pivotal elements have been demonstrated to strongly influence the outcome of BCI training. Ideally, models of Neurofeedback training should capture one or several of the following effects : 
\begin{itemize}
    \item \textbf{F1 : biomarker specificity.} We still have superficial knowledge about brain dynamics and the scientific consensus about how physiological markers relate to a behavioural (and cognitive) features is often approximate and consistently shifting \cite{Picken2020}. Nonspecific biomarkers may be affected by several cognitive dimensions, and regulating them may lead to unwanted cognitive alteration : for instance, motor imagery training \cite{Zhou2022} may cause anxiety \cite{Chen2021} because the chosen biomarkers may share common features.  Training a wide array of possible cognitive states may induce several possible limitations: the cognitive training can fail due to the subject focusing on non-targeted cognitive states that prompt a positive feedback \cite{Mhl2014,Roy2013,Grissmann2017}. Physiological markers depending on several, possibly unrelated cognitive states also constitute poor training indicators, as their value can change depending on state combinations and prevent efficient learning. Finally, the cognitive pathways involved in self-regulating may affect unwanted cognitive dimensions (e.g. "to focus my attention, i have learnt to act in a way that also causes anxiety...").
    \item \textbf{F2 : measurement noise.} The brain signals used to provide a feedback to the subject may be significantly altered during measurement. It may be due to deformation/diffusion through the subject's head, measurement noise sources, loss of resolution, blind spots, etc. The experimenter uses a \textbf{forward model} to reconstruct the raw physiology of the brain using measurements, but these inferences may be flawed and induce significant error sources : for instance, high feedback noise may prevent subjects from effectively finding out which mental strategy worked well to increase the feedback / find patterns in how the feedback responds to their mental actions. This may render the training useless as the indicator seems uncontrollable or random.
    \item \textbf{F3 : user factors (both exogenous and endogenous).} Several factors linked to patient cognition and behaviour have a significant influence on the course of a BCI training. Exogenous elements such as the experimenter instructions before and during the training, the task set or the form of the feedback provided to the subjects have been recognized as significant factors of training success \cite{Jeunet2018,ros_consensus_2020,Roc2021}. The influence of endogenous factors such as subject fatigue, motivation or particular emotions on BCI training have a significant impact on training \cite{Roy2013,Mhl2014,Grissmann2017}, but these effects have also proven difficult to explain. Some modeling studies, such as \cite{oblak_2017}, have made use of spontaneous noise in the brain activity as a way to describe internal endogenous perturbations, but as far as we know, these elements haven't yet been the object of explicit quantitative modelling and we believe it constitutes a key point to explain training outcomes. Thankfully, recent approaches in the field of predictive coding and Active Inference have taken to model the cognitive effect of some endogenous factors (mindfulness, emotions), and provided conceptual frameworks to study their effect on training \cite{hesp_deeply_2021,sandved-smith_towards_2020}.
    \item \textbf{F4 : learning mechanisms.} Because BCI / Neurofeedback training are learning paradigms, they require the experimenter to design a well thought-out environment to guide the evolution of the subject. The nature of the learning mechanisms at play during BCI training has been wildly debated (see above).
    \item \textbf{F5 : placebo effect.} Finally, the impact of placebo effects on the training outcome has proven hard to disentangle from "true" NF effect \cite{Onagawa2023}. The prevailing consensus in neurofeedback studies is to systematically assess Neurofeedback protocols through double-blind studies \cite{fovet_assessing_2017, schabus_better_2017, ros_consensus_2020}. This has led \cite{thibault_neurofeedback_2017} to coin the term of "superplacebo" for NFT : a paradigm in which both experimenter and subject believe in the therapeutic benefits of a practice despite of the lack of supporting evidence, leading to an increased placebo effect and more biaised reporting . Other approaches underlined that the placebo effect was a integral part of the neurofeedback training \cite{Olson2021}, playing a neutralization or amplification role in the training efficacy. Whatever argument eventually prevails, the non-specific mechanisms of NFT related to the technology and the user \cite{Wood2018,Pillette2021}, within or outside of the training proper, are crucial features of theses studies, while still sometimes under-reported. \\
\end{itemize}

\textbf{Existing modeling works :} Modeling approaches have built upon those theoretical leads to create computational representations of neurofeedback training. The objective of these studies is to capture part of the dynamics inherent to neurofeedback training, explain the current experimental results and hopefully predict the impact of some parameters on the training outcome. Previous formalization approaches have focused on defining the key concepts of neurofeedback, most particularly subject feedback perception, feedback control and learning. Gaume et al. \cite{Gaume2016} have proposed a formal description of the psychological dynamics involved in neurofeedback and derived a control-theory inspired model of NFT, featuring both implicit and explicit feedback modalities. However, they did not provide an explicit computational model describing the dynamics of the training . Davelaar has proposed multi-stage,anatomical-based model of EEG brainwave regulation through neurofeedback \cite{davelaar_mechanisms_2018} as well as multi-stage model of neurofeedback subject learning phases \cite{Davelaar2020}. In the former, pools of excitatory and inhibitory neurons were used to generate artificial EEG data, as described in \cite{Izhikevich2003}. An artificial striatal unit made of 1000 binary neurons controlled the peak alpha frequency chosen as marker. The agent looked for the responsible group of neurons through incremental probability updating using the feedback provided to him. This approach was a low-level model of neuron selection which described the physiological pathways involved during self-regulation of the subjects but did not tackle higher-level function. For instance, the model did not describe training effect on user mental state or the impact of prior biases. \cite{oblak_2017} investigated the influence of temporal factors on subjects ability to self-regulate. It proposed a model of subject self-regulation during fMRI neurofeedback training involving a difficult credit assignment problem, modulated by feedback delay and noise. Importantly, two distinct learning structures were compared to test the influence of different subject learning strategies. Cognitive learning was meant to represent explicit strategies the subjects could pick when faced with the feedback, whereas automatic learning was associated with a more implicit modality where the subject tried to control an unconstrained set of cognitive states given a feedback signal, in a formulation reminiscent of reinforcement learning (RL). More recently, a study by Lubianiker and colleagues \cite{Lubianiker2022-ol} featured a family of models similar to the ones shown in this paper, that mapped traditional NFT components to canonical (model-free) RL elements. According to this formulation, the feedback provided to the subjects constitutes states and the NF subjects are assimilated to RL agents having to pick actions by learning a value function, relating those states and potential mental actions. Although the proposed framework constitutes an adequate approximation for the initial interaction between a subject and the BCI loop, it struggles to account for later parts of the training, especially the ability of subjects to learn a more refined representation of the environment or transfer the acquired knowledge in another environment. We argue that a more model-based account of subject experience during NFT is needed to account for these phenomenons and to capture the effect of endogenous factors such as subject motivation, experimenter instructions, and more generally to allow for a better interpretability of training mechanisms and outcomes. \\

\textbf{Motivation :} This paper aims at improving upon the previous approaches to by providing an alternative perspective on BCI/neurofeedback training. All in all, this paper answers a triple need expressed by the neurofeedback community : 
\begin{itemize}
    \item 1. We propose a complete family of models based on the Active Inference framework, allowing us to describe subject learning during NFT, from the initial exploratory steps to the transfer learning. This family of models is generic and may feature a number of alternative hypotheses regarding the driving mechanisms / goals of training.
    \item 2. We show how one may use this formulation to clearly frame NF training goals and hypotheses : \textit{What kind of change is my neurofeedback training aiming for within the subjects ?  What are my assumptions regarding the biomarker I am using ? How do I expect my instructions to affect the training ?}
    \item 3. We show that this family of models may be used to answer researcher questions and provide broad estimators for subject parameters. If the experimenters have precise expectations regarding some key parameters of the training (noise of the biomarker, layout of the cognitive states of the subject, etc.) , we may be able to explain variability in results and eventually predict subject training curves depending on their internal parameters. This may eventually lead to increased therapeutic efficiency through the design of training paradigms tailored for each individual.
\end{itemize}

Our approach breaks from the previously established works by modeling the evolution of subject beliefs during training. Similarly to \cite{Lubianiker2022-ol}, we describe the whole BCI loop as an action selection problem, but we leverage Active Inference to equip the subjects with a broader variety of internal representation(s) of their environment. This allows us to explicitly model subject perceptual uncertainty, the effect of prior beliefs about their environment (beliefs about the feedback, about their ability to control their brain activity), of experimenter instructions, of various mental representations, etc. Contrary to the works of Davelaar and Oblak, our formulation focuses on cognitive aspects of training and does not make explicit the brain activity of the subjects, instead considering a direct mapping between their cognitive activity and the provided feedback. 

\section{Methods}

\subsection{Learning through interaction: a cognitive model of BCI Training} \label{section:neurofeedback_model}

Neurofeedback , and more generally BCI interaction, can be understood as a distinctive form of human–machine interaction characterized by a bidirectional learning process between two primary agents: the subject and the experimenter. In this setting, the subject engages with an artificial environment using modulations of their own neural activity, while the experimenter designs and calibrates the environment's responses to these neural signals. The efficacy of neurofeedback thus depends on the internal models and assumptions held by both parties: the subject's evolving understanding of how to interact with the system, and the experimenter's hypotheses about the neural correlates of cognitive or behavioral traits.\\

\textbf{The experimenter :} When designing the feedback loop, the experimenter typically presumes a relationship between measurable neurophysiological patterns and certain behavioral or cognitive traits of interest—for instance, enhanced theta activity or decreased beta activity in individuals with attentional deficits. Our framework introduces an intermediary conceptual construct—the cognitive system—which serves to link observable behavior with measurable brain activity. Thus, targeting a neural biomarker implicitly entails targeting a constellation of cognitive states that can be externally inferred via behavior. For example, neurofeedback interventions for conditions such as ADHD or insomnia often aim to modulate attentional processes \cite{Arns2020-mb,marzbani_methodological_2016}, under the assumption that these are reflected in specific EEG signal features \cite{EEGCoopeia,Picken2019}.  This cognitive model allows us to decompose the neurofeedback interaction into two interrelated components: (1) a biomarker hypothesis, which posits a mapping between cognitive states and neurophysiological signals, and (2) a forward model, which describes how brain signals are processed and transformed into feedback stimuli. In this light, the feedback presented to the subject constitutes an inferred estimate of their ongoing cognitive state—for instance: \textit{"The feedback is low, which means that your attention level is low"}. \\

\textbf{The subject :} In this environment, subjects have to perform two tasks simultaneously : a perception task and a decision making task. First, they must make sense of the feedback (\textit{"The feedback is low, what does it mean regarding my attention level ?"}). Second, they must figure out a mental strategy that allows them to increase the feedback or control its level voluntarily (\textit{How can my actions change my cognitive state in order to perform well ?}). Note that in this case, the actions performed are mental actions (concentrate, perform mental imagery, recall a memory, inner speech, etc.). To be able to better navigate this environment across training, subjects need to learn the rules of the neurofeedback environment to perform more informed inferences. Our model assumes that the subjects use the provided feedback to build an (approximate) internal representation of his/her environment and learn from an history of feedback observations and mental actions. Note that the complexity of this representation need not match the full neurofeedback loop, but it should allow them to navigate it as efficiently as possible. To be able to capture the effect of intrinsic factors within the NF subjects, like their prior knowledge of self-regulation, our model thus explicitly describes how the beliefs of the subject about their cognitive states evolve across training. \\

To sum-up, we cast NF (and by extension BCI) training as a joint exercise of feedback perception and mental decision-making and learning for the subject. This exercise is particularly arduous due to the numerous uncertainties surrounding the system (see \ref{sec:intro}) and to the numerous factors affecting the training outlined above. Although the nature of the learning mechanisms leveraged during neurofeedback/BCI training is still a matter of debate \cite{enrique_oper_2017,veilahti_neurofeedback_2021}, common machine learning paradigms have been proposed to account for such representation building. The most notorious is probably that of model-free Reinforcment Learning \cite{Lubianiker2022-ol}, though we argue that a more complex model-based approach grounded in the Active Inference framework provides the basis for a broader family of subject representations that may be more applicable in the NF context (e.g., explicit modeling of beliefs/uncertainty, handling partial observability, potential for better generalization/transfer, explaining effect of instructions/priors). In the next section, we introduce this framework and explain how it formalizes subject learning in BCI contexts. \\

For the remainder of this study, we separate an instance of neurofeedback training temporally as follows : the whole training comprises $\mathcal{T}$ specific trials (noted $\dag\in [\![1,\mathcal{T}]\!]$) , which are themselves made of $T$ timesteps. We note $t\in [\![1,T]\!]$ an individual timestep. \\

\subsection{A quick introduction to Active Inference}\label{sec:active_inference}

The Active Inference Framework (AIF) \cite{friston_active_2016,Smith2020} is a broad term that encompasses a family of models able to perform perception, action and learning through the minimization of a single cost function (namely the Variational Free Energy). It provides a biologically plausible framework for modelling subject cognitive states dynamics in response to feedback. This section provides a concise overview of AIF's theoretical foundations and its applicability to representing learning processes in NF and related BCI paradigms.

\subsubsection{Theoretical foundations} \label{section:ai_theory}

\begin{figure}
	\centering
	\includegraphics[width=\linewidth]{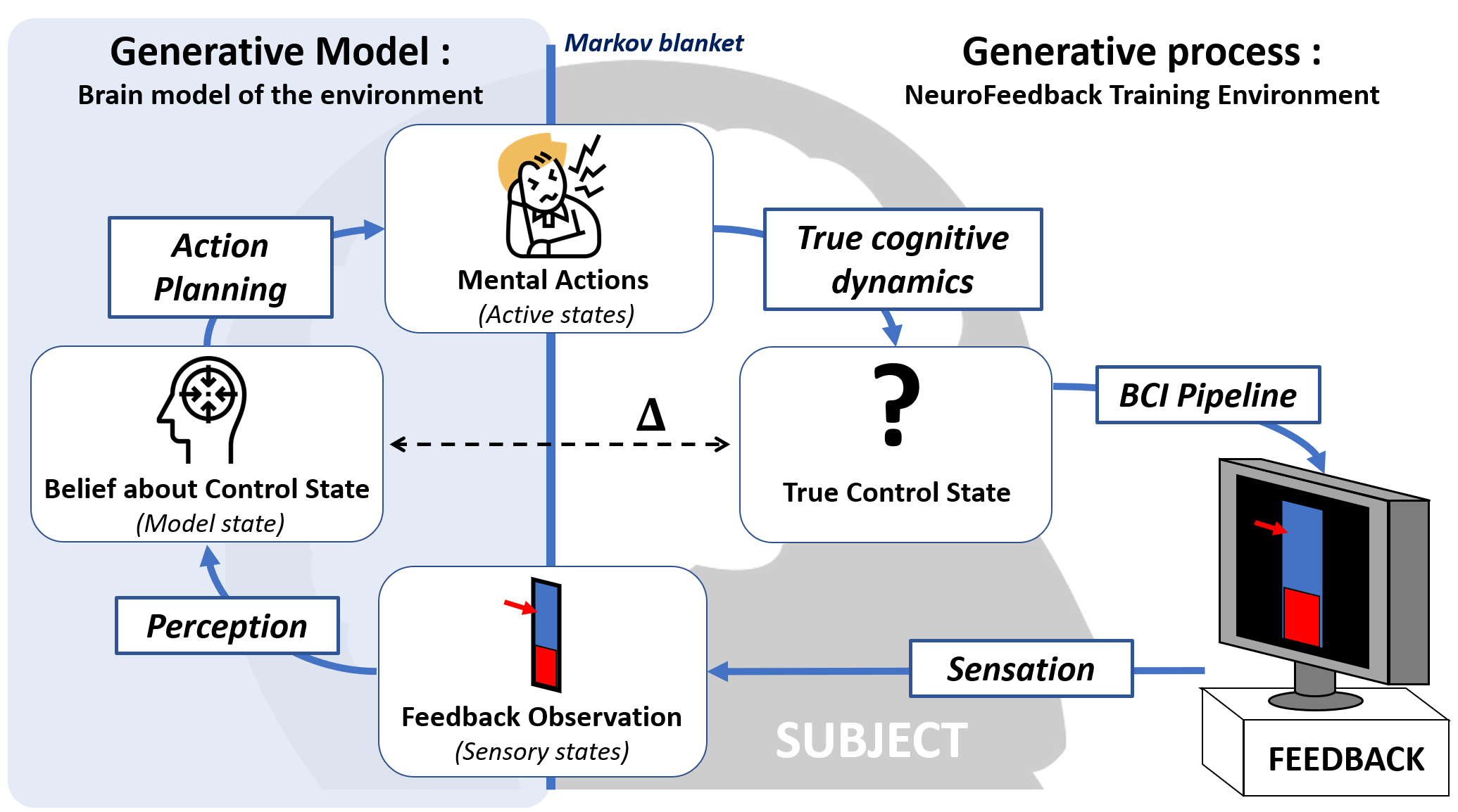} 
	\caption{Generative process and generative model : the subject's brain models the neurofeedback environment to predict outcomes and optimize its actions. Remarkably, BCI training casts cognitive states as external "environment" variables and includes them in its generative model. It is thus necessary to make a distinction between the hidden states themselves and how they are perceived internally by the subject. The agent can only interact with its environment through boundary states (its sensory and active states).}
	\label{fig:feedback_matrices}
\end{figure}

Active Inference operationalizes the Bayesian brain hypothesis \cite{Knill2004}, which posits that the brain constructs and maintains a generative model of its environment to predict incoming sensory data and guide adaptive behaviour. This perspective suggests that maintaining homeostasis (i.e., remaining within viable physiological bounds) requires organisms to continuously minimize prediction errors through updating their internal model and selecting appropriate actions. Perception corresponds to updating beliefs about the current causes of sensory input (model inversion), while learning involves refining the model parameters over longer timescales \cite{Smith2020}.

A key tenet of AIF is the distinction between the generative process—the actual, often unobservable, dynamics of the environment generating sensory data—and the agent's internal generative model—a probabilistic, approximate representation of this process (see Figure \ref{fig:feedback_matrices}). The agent interacts with the external world solely through its sensory inputs (observations) and motor outputs (actions), which constitute the statistical boundary known as a Markov blanket. Consequently, inferring the hidden states of the environment ($s^*$) based on observations ($o$) is fundamentally an inference problem, constrained by the information available at this boundary. The brain must entertain beliefs about the hidden causes of its sensations and update these beliefs in a Bayesian manner. Selecting actions that either reduce uncertainty about the world (exploration) or lead to preferred outcomes (exploitation) allows the agent to maintain accurate beliefs and ensure its continued existence. The inherent limitations on computational resources necessitate a balance between model accuracy and complexity, aligning with principles like Occam's razor, which is implicitly handled by the  VFE minimization \cite{Murray2005}.

AIF formalizes the generative model as a joint probability distribution over observations $o$ and latent variables $\theta$, $P(o,\theta) = P(\theta)P(o|\theta)$ (Eq. \label{eq:gen_model}). Here $\theta$ encompasses hidden states $s$, model parameters, structure  $\mathcal{M}$ and hyper-parameters $\xi$ \cite{da_costa_active_2020}. We assume that observations are caused by those parameters following the \textbf{likelihood} $P(o|\theta)$ . The subject initial belief about the values of those states can be expressed as $P(\theta) = P(s,\xi,\mathcal{M})$ (the so-called \textbf{prior}). To leverage this model, an agent needs to perform inference over the model parameters by computing the posterior distribution $P(\theta|o)$ statistical inference. In effect, it needs to answer the following question: \textit{Given the current observable data, what would be the most likely causes in my model?}. This is done using the famous Bayes' Theorem :

\begin{equation} \label{eq:gen_model_full}
    \underbrace{P(\theta | o)}_{\text{Posterior}} =  \frac{\overbrace{P(o|\theta)}^{\text{Likelihood}}\overbrace{P(\theta)}^{\text{Prior}}}{\underbrace{P(o)}_{\text{Evidence}}}
\end{equation}

However, calculating the normalization constant, the model evidence $P(o) = \int_{\theta \in \Theta} P(o,\theta) d\theta = \int_{\theta \in \Theta} P(o|\theta)P(\theta) d\theta$ (equation \label{eq:calc_evidence}) is typically computationally intractable for complex models due to the high-dimensional integration involved. \textit{As an example of the previous assertion, let's imagine, for a fixed model structure $m_0$ and a given set of hyper-parameters $\xi_0$, a set of 3 latent states $s_{1,2,3}$ with 10 possible values each (which is arguably a very simple environment). Such a state space leads to a summation over 1000 values of $\theta$ to compute an individual posterior $P(\theta_i | o)$, and therefore 1000 computations of likelihoods $P(o|\theta)$.} In practice, this vulnerability to the curse of dimensionality has led to a need to either shunt the calculation of $P(o)$ (as in Maximum-A-Posteriori estimators) or approximate $P(o)$ using methods like Monte-Carlo Sampling. \\

To circumvent this intractability, AIF employs Variational Inference (VI). VI reframes the inference problem as an optimization problem by introducing an approximate posterior distribution $q(\theta|\chi)$ parametrized by variational parameters $\chi$. The aim of variational inference is to minimize a measure of distance between the true distribution $P(\theta|o)$ and our proposed approximation. Mathematically, we can define the optimal approximation as the function minimizing the KL-divergence between $P$ and $q$ :

\begin{align*}
    q^* = q(\theta|\chi^*) \text{ with } \chi^* &= \argmin_{\chi} D_{KL}[q(\theta|\chi)||P(\theta|o)]\\
    &=  \argmin_{\chi} \int_{\theta\in \Theta} q(\theta |\chi)\ln\frac{q(\theta |\chi)}{p(\theta |o)}d\theta \text{\textit{   (Continuous formulation)}} \\
    &=  \argmin_{\chi} \sum_{\theta\in \Theta} q(\theta |\chi)\ln\frac{q(\theta |\chi)}{p(\theta |o)} \text{\textit{    (Discrete formulation)}}
\end{align*}

\textit { Where $D_{KL}$ is the Kullback-Leibler Divergence between two distributions. If $D_{KL}[Q||P] = 0$, then $ Q = P$ and the inference problem is solved.} Effectively, it means that the previous inference problem, has been cast as a more 'classical' optimization problem. \cite{millidge2021} Further transformations give :

\begin{equation}
    \label{eq:err_to_vfe}
    \begin{aligned}
        \underbrace{D_{KL}[q(\theta|\chi)||P(\theta|o)]}_{\text{Error to minimize}} &= D_{KL}[q(\theta|\chi)||\frac{P(\theta,o)}{P(o)}]\\
        &= D_{KL}[q(\theta|\chi)||P(\theta,o)] + \mathbb{E}_{q(\theta|\chi)}[\ln P(o)]\\
        &= \underbrace{D_{KL}[q(\theta||\chi)||P(\theta,o)]}_{\text{Variational Free Energy}} + \underbrace{\ln P(o)}_{\text{ - Surprise }}\\
    \end{aligned}
\end{equation}

We can thus define a quantity $\mathcal{F} = D_{KL}[q(\theta|\chi)||P(\theta,o)]$ which verifies $\mathcal{F} \geq D_{KL}[q(\theta|\chi)||P(\theta|o)]$. $\mathcal{F}$ is the \textbf{Variational Free Energy}, which creates an upper bound on the error between the true posterior and the proposed approximation. The negative free energy term $-\mathcal{F}$ also provides a lower bound on model (log) evidence (ELBO). Maximizing the ELBO / minimizing the variational free energy would allow us to minimize "surprisal", which is defined as the negative log probability of a given outcome $-\ln P(o)$ (not to be confused with psychological 'surprise'). Variational inference consists in minimizing Variational Free Energy (VFE) which mechanistically minimizes two quantities: the difference between the true posterior distribution $P(\theta|o)$ and its approximation $q$ as well as surprisal, in effect maximizing model evidence. For the next steps, we assume a discrete observation and latent state space.\\

\begin{align} \label{eq:vfe}
    \mathcal{F} &= \sum_{\theta \in \Theta} q(\theta|o,\chi)\ln\frac{q(\theta|o,\chi)}{P(o,\theta)} = \mathbb{E}_{q(\theta|\chi)}[\ln\frac{q(\theta|\chi)}{P(o,\theta)}] - \ln P(o)
\end{align}

By minimizing the free energy functional expressed in \ref{eq:vfe}, agents thus build increasingly accurate models of their environment. \\

\textbf{Planning as inference:} AIF extends this minimization principle to action selection. Agents are assumed to select actions, or sequences of actions (policies  $\pi$), in order to minimize the Expected Free Energy (EFE) under a policy \cite{sajid_active_2021,friston_active_2017}. EFE quantifies the free energy expected upon executing a particular policy, considering future potential observations and state transitions. By rewriting the heavy notation $q(\theta|\chi)$ to $q(\theta|\pi)$, we get \cite{sajid_active_2021} : 

\begin{align} \label{eq:vfe_policy}
	\mathcal{F_\pi} &= \underbrace{D_{KL}[q(\theta|\pi)||P(\theta|o,\pi)]}_{\text{Evidence bound}} - \underbrace{\ln P(o)}_{\text{ Log evidence }}
\end{align}

Planning in such a way naturally balances the expected utility of outcomes (reaching preferred states, encoded in prior preferences over outcomes) and the expected information gain (reducing uncertainty about the hidden states or model parameters) \cite{parr_generalised_2019}. Our implementation utilizes an expansion of the original Active Inference planning scheme : the sophisticated inference scheme \cite{sophisticated_inference}, which performs a tree search over possible future action-outcome sequences to evaluate EFE and select the optimal policy, albeit potentially computationally intensive for long planning horizons. 

This unified formulation allows agents to build increasingly coherent internal generative models, be it through inferring the hidden variables in the model, acting to resolve uncertainty about them or learning the intrinsic dynamics of the system. In essence, AIF posits that perception, action selection, and learning all emerge from a single imperative: minimizing free energy (VFE for perception and learning, EFE for action selection). This happens accross three different steps: 1) during inference, the most likely hidden states are guessed using knowledge about hidden state evolution and observations. This is a very short-term mechanic, which is leveraged once per timestep during sequential simulations. 2) After state inference, agents perform actions to pursue desired outcomes \cite{Smith2020} and resolve uncertainty. 3) Finally, the agent changes its general model of the world. This evolution affects wider timescales by updating its representation of general environment dynamics. Together, these three instrumental mechanisms provide a general description of agent planning and behaviour during a single trial as well as accross a large number of trials : 

With $\theta = (\underbrace{\theta_{s_{t\in[0,T]}}}_{\text{Belief about latent states}},\overbrace{\theta_{u_{t\in[0,T-1]}}}^{\text{Agent actions / policy }},\underbrace{\theta_{\alpha}}_{\text{Agent graph hyperparameters (see next section)}})$ :

\begin{subequations}
    \label{eq:free_energy_components}
    \begin{empheq}[left={\empheqlbrace\,}]{align}
        \theta_{s_t} &= \argmin_{\theta_{s_t}} \mathcal{F}(\theta,o_t)\label{eq:state_inference_fe}\\
        \theta_{u_t} &= \argmin_{\theta_{u_t}} \mathcal{F}(\theta,o_t) \label{eq:action_inference_fe} \\
        \theta_{\alpha} &= \argmin_{\theta_{\alpha}} \sum_{t = 0}^{T} \mathcal{F}(\theta,o_t) \label{eq:learning_inference_fe}
    \end{empheq}
\end{subequations}

With \ref{eq:state_inference_fe} the hidden state inference (computed every timestep), \ref{eq:action_inference_fe} the action selection (computed every timestep) and \ref{eq:learning_inference_fe} the environment dynamics learning (computed at the end of one trial).

\subsubsection{Discrete Active Inference : Variational Inference in a Markov Decision Process}\label{sec:active_inference_learning}

\begin{figure}
    \includegraphics[width=\linewidth]{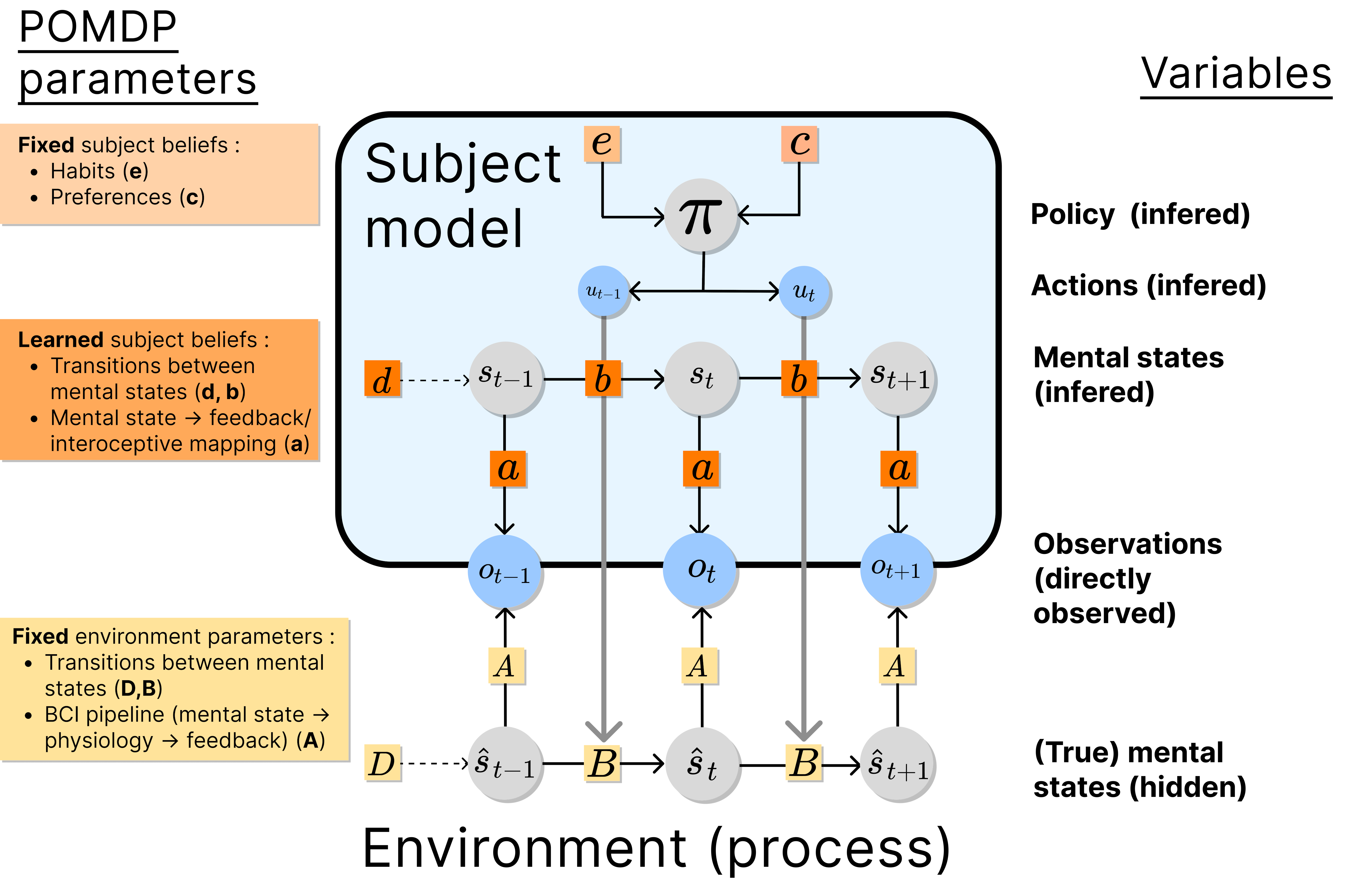}
    \caption{The Partially Observable Markov Decision Process (POMDP) used to model training. The environment (generative) process figures a set of hidden states ($\mathbf{\hat{s}_t}$), corresponding to the subject "actual" cognitive states in our formulation. Although impossible to see directly, each state stochastically generates stimuli observable by the subject ($o_t$) depending on a true observation function $\mathbf{A}$. It may consist in some external feedback or in an interoceptive observer the subject must learn to interpret. Possible state transitions ($\mathbf{B}$) and starting states ($\mathbf{D}$) are fixed before training. The subject generative model of the cognitive regulation is shown above, featuring a set of perceived states  ($\mathbf{s_t}$). The subject models the feedback as a realization of hidden states with the function $\mathbf{a}$, and the effect of his/her mental actions $u_t$ on those states with the function $\mathbf{b}$. The Active Inference agent uses this formulation to update their model on two distinct timescales by minimizing their Free Energy following the equations described in \cite{sophisticated_inference,da_costa_active_2020}. On the timestep timescale, it infers the hidden states that best match observations and prior beliefs, and the actions that best match its habits/preferences/exploration drive. On the trial timescale, the agent updates its beliefs about the environment dynamics ($\mathbf{a}$,$\mathbf{b}$,$\mathbf{d}$) to better predict and navigate it.} 
    \label{fig:ai_process}
\end{figure}

To apply AIF to NF training, we model the subject-environment interaction using a discrete Partially Observable Markov Decision Process (POMDP), as depicted in Figure \ref{fig:ai_process}. This formalism allows us to explicitly differentiate between the true environmental dynamics (generative process) and the subject's internal representation (generative model).

The generative process defines the objective reality of the NF loop. It is formalized using a Hidden Markov Model which explicitely describes the true latent cognitive state of the subject $\hat{s_t}$, as well as the rules of the environment : 
\begin{itemize}
	\item The true distribution of initial states $P(\hat{s}_0)$ parametrized by the upper-case vector $\mathbf{D}$.
	\item The true transitions between states influenced by subject actions $P(\hat{s_{t+1}}|\hat{s_t},u_t)$ parametrized by the upper-case matrix $\mathbf{B}$.
	\item The true mapping from these states to observable feedback ($o_t$), $P(o_t|\hat{s_t})$ parametrized by the upper-case matrix $\mathbf{A}$.
\end{itemize}
These components proposed a description of the subject brain activity and feedback design and remained fixed throughout a simulation.

The generative model represents the subject's subjective beliefs and understanding of the NF environment. Contrarily to the generative process, the generative model uses categorical probability distribution to represent the current (belief) state, denoted using the bold-case $\mathbf{s_t}$. The generative model includes the following mappings : 
\begin{itemize}
	\item The subject's beliefs about the initial state distribution, $P(s_0)$ parametrized by the lower-case vector $\mathbf{d}$.
	\item The subject's beliefs about the state transition rule $P(s_{t+1}|s_t,u_t)$ parametrized by the lower-case matrix $\mathbf{b}$.
	\item The subject's beliefs about the observation mapping, $P(o_t|s_t)$ parametrized by the lower-case matrix $\mathbf{a}$.
	\item The subject's prior preferences towards certain observations, $P(o_t)$ parametrized by the lower-case vector $\mathbf{c}$.
	\item The subject's prior over actions, representing biaises towards recurring strategies , $P(u_t)$ parametrized by the lower-case vector $\mathbf{e}$.
\end{itemize}
These matrices have initial values that encode subject prior beliefs about their environments (e.g. \textit{before training, how do I think the feedback relates to my hidden cognitive state ?}), but may evolve significantly through learning across trials depending on observed outcomes (feedback levels, performed actions). This representation allows the agent to maintain an easily interpretable model of its environment, and a way to quantify how confident the agent is about the dynamics it models.

\subsubsection{Mathematical Formalism}\label{sec:math_heavy}

This section details the core mathematical equations governing perception, planning, and learning within the AIF-POMDP framework. We assume the subject potentially perceives multiple observation modalities ($m$) arising from multiple hidden state factors ($f$). Vectorized parameters are bolded. For simplicity, we omit factor/modality superscripts where unambiguous.

\textbf{Emission mappings:} (formally $P(o_t|s_t)$).

For the generative process :
\begin{align}
    \text{For } t \in [\![0,T]\!],\mathbf{o_t} &\thicksim \mathbf{A} [\hat{s}_t]
\end{align}

For the generative model :
\begin{align}
	\text{For } t \in [\![0,T]\!],P(o_t|\mathbf{s_t},\mathbf{a}) &= Cat(\mathbf{a}_N \mathbf{s_t})
\end{align}

\textit{Where $\mathbf{a}_N$ is the normalized $\mathbf{a}$ matrix to ensure it sums to 1 for each state $s_t$, $\thicksim$ is the sampling operation and Cat is the categorical distribution.} \\

\textbf{Initial states and transitions:} (formally $P(s_0)$ and $P(s_{t+1}|s_t,u_t)$) : 

For the generative process :
\begin{subequations}\label{eq:process_transitions}
	\begin{empheq}[left={\empheqlbrace\,}]{align}
		&\qquad \hat{s}_{0} \thicksim \mathbf{D} \label{eq:ai_d} \\
		\text{For } t \in [\![0,T-1]\!], u\in [\![1,U]\!],&\qquad \hat{s}_{t+1} \thicksim \mathbf{B}_u [\hat{s}_t] \label{eq:ai_b}
	\end{empheq}
\end{subequations} 

For the generative model : 
\begin{subequations}\label{eq:model_transitions}
	\begin{empheq}[left={\empheqlbrace\,}]{align}
		&\qquad P(s_{0}|\mathbf{d}) = Cat(\mathbf{d}) \\
		\text{For } t \in [\![0,T-1]\!], u\in [\![1,U]\!],&\qquad P(s_{t+1}|\mathbf{s_t},u_t,\mathbf{b})= Cat(\mathbf{b}_N[u_t] \mathbf{s_t})
	\end{empheq}
\end{subequations} 
\textit{Where $\mathbf{b}_N$ is the normalized $\mathbf{b}$ matrix to ensure it sums to 1 for each state / action pairs $s_t,u_t$.}\\

\textbf{Perception :} Under the sophisticated inference scheme, the posterior belief update simplifies to a combination of likelihood evidence and prior beliefs based on the previous state and action, implementing a practical Bayesian filter :
\begin{equation}
    \label{eq:perception_update}
    \begin{aligned}
        \mathbf{s_t} = \sigma(ln(\mathbf{a}_N \cdot \mathbf{o_t}) + ln(\mathbf{b}_N[u_t] \mathbf{s_{t-1}}))
    \end{aligned}
\end{equation}
\textit{Where $\sigma$ is the softmax function and $\cdot$ is the inner product, meaning that $\mathbf{a}_N \cdot \mathbf{o_t} = \mathbf{a}_N^T\mathbf{o_t}$}.\\

\textbf{Planning and decision-making :} Active inference agents attempt to minimize the \textit{expected} free energy \cite{friston_active_2017} (EFE) of their model by performing actions that (a.) lead to preferred outcomes (in our case, a high positive feedback level) and (b.) improve the agent knowledge about its environment. The EFE can be seen as a sort of objective function when the agent plans its next move. Sophisticated inference agents \cite{sophisticated_inference} build a tree of future actions and outcomes to plan their next actions. This planning scheme was chosen over traditional active inference because it allowed for more flexible action selection without requiring predefined action sequences (or policies). For each prospective timestep $\tau$, an action - outcome branch $(o_t,u_t)$ has the following negative EFE :
\begin{equation}
    \label{eq:efe_1}
    \begin{aligned}
        G(u_{\tau},o_{\tau}) &= \underbrace{g_u(\mathbf{s_{\tau+1}^u},\mathbf{s_{\tau}^u},\mathbf{o_{\tau+1}^u})}_{\text{EFE for the timestep $\tau+1$}} + \underbrace{\mathbf{u_{\tau+1}^o}.G(u_{\tau+1},o_{\tau+1})\mathbf{o_{\tau+1}^u}}_{\text{EFE for the subsequent timesteps}}
    \end{aligned}
\end{equation}
$g_u(\mathbf{s_{\tau+1}^u},\mathbf{s_{\tau}^u},\mathbf{o_{\tau+1}^u})$ is the (negative) EFE estimator for the prospective timestep $\tau+1$ (and only this one !) if the expected state and observations after action $u_{\tau}$ and observation $o_{\tau}$ are $(\mathbf{s_{\tau+1}^u},\mathbf{o_{\tau+1}^u})$ and the previous posterior $\mathbf{s_{\tau}^u}$. In this study, we used the following (negative) EFE estimator : 

\begin{equation}
    \label{eq:efe_2}
    \begin{aligned}
        g_u(\mathbf{s},\mathbf{s_-},\mathbf{o}) &= \underbrace{\mathbf{e}[u]}_{\text{Habits}} + \underbrace{\mathbf{o}[ln \mathbf{o} + \mathbf{c}]}_{\text{Risk}} + \underbrace{\mathbf{s} . \mathbf{H}}_{\text{Ambiguity}} - \underbrace{\mathbf{o}\cdot\mathbf{W_a}\mathbf{s}}_{\substack{\text{Novelty-seeking} \\ \text{(emissions)}}} - \underbrace{\mathbf{s}\cdot\mathbf{W_b}[u]\mathbf{s_-}}_{\substack{\text{Novelty-seeking} \\ \text{(transitions)}}} \\
        \mathbf{H} &= -diag(\mathbf{a} \cdot ln\mathbf{a}) \\
        \mathbf{W_a} &= \frac{1}{2}(\mathbf{a}^{\odot-1}-\mathbf{a}_0^{\odot-1})\\
        \mathbf{W_b} &= \frac{1}{2}(\mathbf{b}^{\odot-1}-\mathbf{b}_0^{\odot-1})
    \end{aligned}
\end{equation}
\textit{Where $\textbf{H}$ is the entropy vector associated with $\mathbf{a}$ and $\mathbf{W_a}$ and $\mathbf{W_b}$ are novelty terms computed every timestep pushing subjects towards prospects of yet unexplored state transitions or outcome generation. $\mathbf{X}_0$ is the sum of all the Dirichlet counts for a parameter matrix $\mathbf{X}$ and $\mathbf{X}^{\odot-1}$ is the element-wise inverse of all elements of matrix $\mathbf{X}$}.

This results in a counterfactual planning scheme in which the agents visualize every plausible future action-outcome paths and seek the ones who maximize the feedback preference (risk term) but also allow agents to improve their knowledge of the environment (ambiguity and novelty seeking terms). Finally, mental actions are picked through sampling the one resulting in the smallest EFE : 
\begin{equation}
    \label{eq:efe_3}
    \begin{aligned}
        u_t \thicksim \sigma(\gamma G(u_t,o_t))
    \end{aligned}
\end{equation}
\textit{With $\gamma$ the action selection inverse temperature, leading to more random action selection when $\gamma \xrightarrow{} 0$ and deterministic action selection when $\gamma \xrightarrow{} +\infty$.}

The Expected Free Energy (EFE) estimator presented here can be modified to account for agents who are less novelty-driven by removing the weights $\mathbf{W_a}$ or $\mathbf{W_b}$ or changing the prior knowledge of the agents (accounting for prior instructions or beliefs about the dynamics of the neurofeedback system).In practice, the tree search demonstrated was conducted for short temporal horizons, which is not an issue as we assumed a gradual preference matrix for the subjects.\\

\textbf{Learning:} agents use Bayesian learning to update their representation of latent environment variables $\mathbf{a},\mathbf{b},\mathbf{d}$ through a correlation-based incremental process. After a trial, a subject uses previous observations and inferences to update its model through Bayesian belief updating. Because the latter is generally hard to compute (for dimensional-related reasons explained before), Active Inference uses the Dirichlet distributions as the conjugate prior for the categorical distribution :

\begin{subequations}
    \label{eq:weights_update}
    \begin{empheq}[left={\empheqlbrace\,}]{align}
        \mathbf{d}_{\dag+1} &= \mathbf{d}_{\dag} + \mathbf{s}_{\dag,0}\\
        \mathbf{b}_{\dag+1} &= \mathbf{b}_{\dag} + \sum_{t=0}^{T-1} \mathbf{s}_{\dag,t+1} \otimes \mathbf{s}_{\dag,t} \otimes \mathbf{u}_{\dag,t}\\
        \mathbf{a}_{\dag+1} &= \mathbf{a}_{\dag} + \sum_{t=0}^{T} \mathbf{o}_{\dag,t} \otimes \mathbf{s}_{\dag,t}
    \end{empheq}
\end{subequations}
\textit{Where $\otimes$ is the outer tensor product and $\forall t,\dag$, $\mathbf{s}_{\dag,t}$ is the state posterior at time $t$ computed during trial $\dag$.} \\
This means that the learned posterior is also a Dirichlet distribution, and computing the posterior given new data is made very easy \cite{KuleshovErmon_BayesianLearning}. In essence, the Active Inference agent counts specific state/observations co-occurrences and increments the Dirichlet parameters of its model (the $a,b,d$ distributions) accordingly. This allows agents to learn flexible representations of their environment.\\

This learning mechanism operates on a slower timescale (trial-by-trial) compared to perception and action selection (timestep-by-timestep). The duration of trials can influence initial exploration patterns. Due to the accumulative nature of Dirichlet updates, prior beliefs become increasingly entrenched with experience, potentially reducing flexibility in adapting to changes unless mechanisms such as parameter decay ('forgetting') are introduced \cite{hesp_deeply_2021}. Importantly, effective learning requires informative priors; agents with uninformative (e.g., flat) initial priors struggle to build coherent models from observations \cite{ma2022}, underscoring the potential role of instructions or prior experiences in facilitating BCI skill acquisition.

Numerous approaches \cite{friston_active_2016,Parr2019,sophisticated_inference,Tschantz2020,da_costa_active_2020,hesp_deeply_2021,sandved-smith_towards_2020,smith_step-by-step_2021} have documented how Variational Free Energy and Expected Free Energy gradients are computed with respect to subject model variables and parameters during state inference, action inference and learning. In this section, we have provided a simplified account of the model updates but we strongly encourage the interested reader to see the previously cited approaches for a more complete explanation of the computations behind our simulations.\\

\subsection{Neurofeedback training modeling with Active Inference}\label{sec:active_inference_nf}

\begin{table}[!ht]
    \begin{center}
    \centerline{
        \begin{tabular}{||m{25em} | m{25em}||}
            \hline
            \textbf{Neurofeedback training components} & \textbf{Active Inference Component} \\
            \hline \hline
            The latent cognitive activity of the subject, driving the feedback through a mapping with brain signals & Hidden states vector $\mathbf{\hat{s}}$. \\
            \hline
            The subject perception of its own cognitive activity. \textit{In a first order approximation, we consider that this perceptive state has negligible influence on the biomarker.}  & Perceived states vector $\mathbf{s}$. \\
            \hline
            Feedback provided to the subject & Observation vector $\mathbf{o}$ \\
            \hline
            \textbf{True} relationship between subject cognitive activity and perceptible feedback. Affected by noise, biomarker hypothesis, experimenter choices regarding pipeline design, etc. & True perception matrix $Cat(\mathbf{A})=P(o|\hat{s})$. \\
            \hline
            \textbf{Perceived} relationship between subject cognitive activity and perceived feedback. Affected by experimenter instructions regarding feedback meaning, belief about feedback efficacy, blindness of patients regarding treatment, etc. & Perception model $Cat(\mathbf{a})=P(o|s)$. \\
            \hline
            \textbf{True} dynamics of cognition. Effect of subject mental actions and spontaneous evolution. & True transition/action matrix $Cat(\mathbf{B})=P(\hat{s}_{t+1}|\hat{s}_{t},u_t)$ (and initial states $Cat(\mathbf{D})=P(\hat{s}_0)$). \\
            \hline
            \textbf{Perceived} dynamics of cognition. Subject representation of its own mental actions and ability (or inability) to control its own states. Affected by experimenter instructions regarding mental strategies to control the feedback, previous knowledge, etc. & Transition model $Cat(\mathbf{b})=P(s_{t+1}|s_{t},u_t)$ (and initial states model $Cat(\mathbf{d})=P(s_0)$). \\
            \hline
            Subject endogenous drive towards high performances (cost of effort, expectations behind treatment, motivation, fatigue, etc.) & Preference matrix $Cat(\mathbf{c})= P(o_t)$. \\
            \hline
            Subject bias towards a certain mental strategy & Subject habits $Cat(\mathbf{e}) = P(u_t)$. \\
            \hline
            Simulate control groups with various types of feedback (sham, feedback of varying bias and noise) and participants & Subjects with varying priors ($\mathbf{a}$,$\mathbf{b}$,$\mathbf{c}$,$\mathbf{d}$ and $\mathbf{e}$) or experiments with varying characteristics ($\mathbf{A}$,$\mathbf{B}$,$\mathbf{D}$,etc.). \\
            \hline
            Report subjects training curves. & Simulated agents cognitive states $s$ and feedback levels $o$ reached across trials. \\
            \hline
        \end{tabular}
    }
    \end{center}
    \caption{Correspondence between traditional components of NeuroFeedback Training systems (from \cite{ros_consensus_2020}) and the Active Inference generic components we use to simulate training.}
    \label{tab:nf_comp_ai}
\end{table}

\begin{figure}[!ht]
    \begin{subfigure}{\textwidth}
        \includegraphics[width=\textwidth]{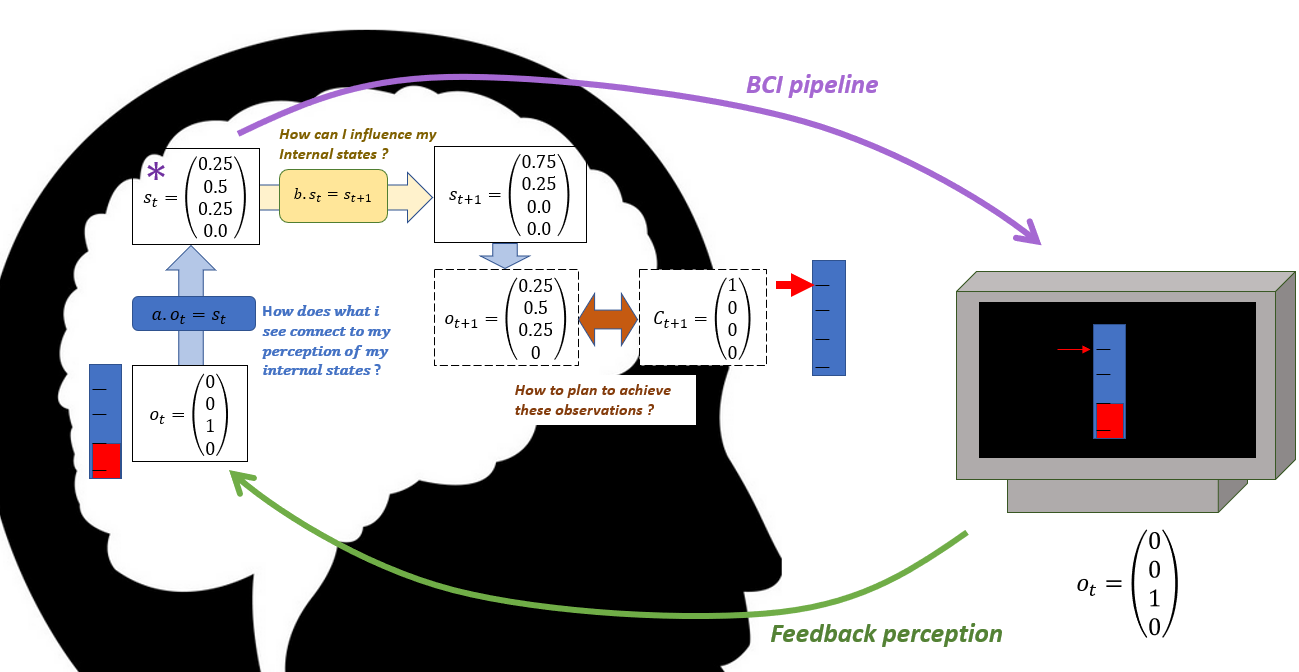} 
            \caption{\textbf{Inference during trials : } The subject uses the feedback (right) to infer the true hidden mental state $s_t$ (marked with a star) given its model of the world (matrices $a$ for perception and $b$ for transitions). Planning then occurs when the subject infers his/her next action to get to specific hidden states $s_{t+1}$ and achieve prefered observations (see matrix $C$). Note that state inference and action selection are complex processes accounting for various MDP dynamics and balancing information-seeking and reward-seeking behaviour in an all encompassing (Expected) Free Energy minimization dynamic \cite{Smith2022}.}
        \label{subjective_models:1a}
    \end{subfigure}

    \begin{subfigure}{\textwidth}
        \includegraphics[width=\textwidth]{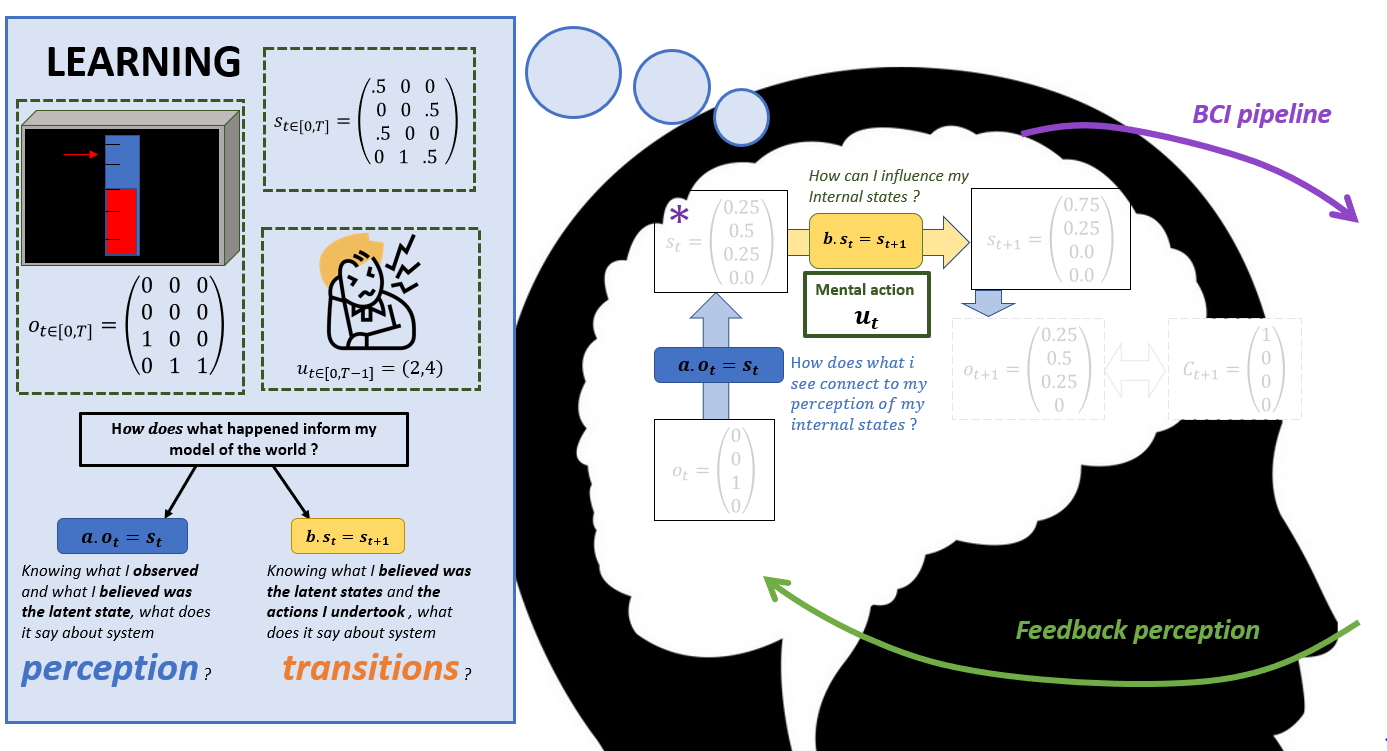} 
            \caption{\textbf{Learning between trials : } On a slower timescale, the agent uses its belief about what the hidden states were during the previous trial $s_{t\in}[0,T]$ to update its model of the environment through correlations. The history of the observations are used to update the perception model $a$ and the history of its actions $u$ to update the transition model.}
        \label{subjective_models:1c}
    \end{subfigure}
    
    \caption{The Active Inference framework unifies agent perception, action and learning by iteratively minimizing these functions across different timescales.}
    \label{fig:active_inference_nf_sumup}
\end{figure}

\subsubsection{Models of training under uncertainty}

Applying the AIF-POMDP framework allows us to formally model the subject's experience and learning trajectory during NF training, explicitly addressing the inherent uncertainties involved. Table \ref{tab:nf_comp_ai} provides a direct mapping between conventional NF components, drawing from consensus guidelines \cite{ros_consensus_2020}, and the corresponding elements within our AIF model.

NF / BCI training is quite distinct from most behavioural paradigms usually explored with the AIF Framework. One of the major distinction rests in the fact that the paradigm is affected by a very significant amount of uncertainty. Two primary sources of uncertainty are central to the NF learning challenge and are captured naturally within AIF: 1) the \textbf{Perceptual Uncertainty} : the ambiguity regarding the relationship between the subject's latent cognitive state and the observed feedback signal. This is represented by the discrepancy between the true likelihood mapping ($\mathbf{A}$) and the subject's learned model ($\mathbf{a}$). Factors like signal noise, the choice of biomarker, and feedback processing contribute to $\mathbf{A}$ whereas experimenter instructions and prior beliefs shape the initial $\mathbf{a}$. 2) \textbf{Control Uncertainty:} Ambiguity regarding the topology of the relevant cognitive state space and the effectiveness of specific mental actions in transitioning between states. This is represented by the discrepancy between the true state transition dynamics ($\mathbf{B}$) and the subject learnt model ($\mathbf{b}$).\\

We use the canonical Active Inference model components to describe subject experience during a mental training task. This entails describing how the subject cognitive activity evolves in response to his / her perception of the feedback and, in a second time, how a subject may select a variety of mental actions that affect it. We formalize the cognitive topology explored by the subjects as a set of discrete latent states $\mathbf{\hat{s}}$. Not all states are modelled equal however! Some are more preferable than others because they correlate with experimenter preferences regarding subject brain signals. Because cognitive dynamics are driven by Bayesian dynamics in the Active Inference formulation, the way beliefs evolve when confronted with new observations is heavily influenced by the quality of initial priors and by the level of subjective uncertainty the subjects entertains. This idea is central to the way we model neurofeedback and its therapeutic efficacy.Figure \ref{fig:active_inference_nf_sumup} provides a visual representation describing how we employ the overarching Active Inference framework to elucidate subjects' perception, planning, and learning throughout various BCI training sessions. 

\textbf{Mental actions:} The agent can change its current cognitive state $\hat{s}_t$ by performing mental actions $u_t$. The agent may not initially know how its action will influence its mental states. Thus it needs a transition model of its own mental actions $\mathbf{b} = P(s_{t+1}|s_t,u_t)$. In practice, we use a matrix to keep track of this model. The true dynamics of cognitive regulations, written $\mathbf{B} = P(\hat{s}_{t+1}|\hat{s}_t,u_t)$ are unknown by the subject and must be inferred.

\textbf{Perception:} The agent is not able to directly observe what its current mental state is, but must infer it. We represent this belief as a distribution over the current states $\mathbf{s}$. To do that, it needs to rely on external (feedback) or internal (meta-cognitive or interoceptive) observation modalities. Again, the agent may not know the true mapping $\mathbf{A} = P(o_{t}^(1),...,o_{t}^(M)|\hat{s}_t)$ and tries to approximate it by updating its own distribution $\mathbf{a} = P(o_{t}^(1),...,o_{t}^(1)|s_t)$. (here, $1,...,M$ are the various observation modalities. In most BCI paradigms, 1 feedback modality is provided but there are exception, notably multimodal feedback and interoception)

Thankfully, one does not need a perfect model of the feedback to derive evidence in order to learn how to control it. Humans have been shown able to derive coherent ans structured models of interactions from broad priors. However, in the precise case of non-invasive BCI interaction, the low signal-to-noise ratio adds a layer of complexity to the already tough inference problem and may explain poor subject performances. As is often the case in natural environments, the agent generative model and the process it tries to mimic may end up differing wildly and lead to inaccurate hidden state perception and sub-optimal actions. A subject model's quality is not evalutated on its ability to mirror the causes of its observations perfectly (which would be both incredibly complex and computationally implausible), but on its capacity to support accurate predictions and goal-directed control.\\

In considering this representation of the BCI training loop, one may pose the following question: \textit{according to this formulation, what would be the mechanistic goal of NF / BCI training ?} This representation proposes several answers :  First and foremost, learning within the model corresponds to the agent progressively mapping its internal cognitive landscape and discovering effective mental actions for navigation, while potentially refining its model of the proposed (BCI-induced) feedback signal. Then, the reinforcement of action sequences leading to desired feedback results in the formation of robust control strategies, analogous to habit formation. Finally, the standard neurofeedback loop may be expanded to incorporate an interoceptive observation modality. Under this formulation, the subject receives two types of sensory information at each timestep: the explicit, externally provided feedback, and an internal, interoceptive signal. Crucially, learning in this context involves not only interpreting the explicit feedback but also developing the ability to recognize and regulate internal bodily signals associated with successful control. This has been cited as a potential enabler of BCI training and a key mechanism in ensuring post-training homeostasis \cite{davelaar_mechanisms_2018}.

\subsubsection{Underlying hypotheses}

Our Active Inference formulation of BCI cognitive training makes a few implicit modeling hypotheses. First, and most importantly, we dissociate subject perception of their own cognitive state (\textit{"What I believe about my current attention level"}) and their actual cognitive state (\textit{"My current attention level"}). For example, in the case of attention training, we assume that the subject's cognitive attention level is distinctly regulated by an independent process during a neurofeedback session. This assumes that the subject needs to make an inference on its current attention state by using the feedback signal and potential internal sensations. Moreover, we assume that the feedback signal is not directly affected by the subject regulatory process. 

Second, it assumes the structure of cognitive state space of the regulated dimension as well as how the subject represents it. To propose an accurate description of neurofeedback training, we have to set a plausible space of cognitive states the agents may navigate to mimic training exercises. Due to the uncertainty surrounding most mental training paradigms however, it is necessarily an approximation. In this paper, we propose a very generic graph in order to showcase the capacities of the leveraged framework in modelling different neurofeedback properties. We discuss this modelling choice and its implications in the discussion.

Third, the tabular Active Inference formalism discretizes the feedback values, the topographies of (true and perceived) cognitive states as well as the actions a subject may select. Although continuous spectra may be more sensible representation for some of those model variables (feedback or actions for instance), making use of Active Inference's discrete formulation allows us to create more complex cognitive graphs and interactions, while retaining computational tractability.

\subsubsection{Neurofeedback Inference Problem and performance indicators}

Within this framework, the agent's task during each NF trial $\dag \in [\![1,\mathcal{T}]\!]$ can be cast as an inference problem over the sequence of states, actions, and model parameters that best explain the observed feedback under the agent's generative model. Formally, the agent seeks to optimize its beliefs about:
\begin{equation}
	\label{eq:nf_inf_pb}
	\tag{NeuroFeedback Inference problem}
	\begin{aligned}
		\theta_{NF}^{\dag} &= (\underbrace{\theta_s}_{\text{Hidden states}}, \overbrace{\theta_u}^{\text{Policy selection}} , \underbrace{\theta_{\alpha}}_{\text{Model parameters}}) \\
		&= (\theta_{s_{0:\mathcal{T}}}, \theta_{u_{0:\mathcal{T}-1}}, \theta_{a}, \theta_{b})
	\end{aligned}
\end{equation}
where $\theta_{s_{0:\mathcal{T}}}$ represents beliefs about the sequence of hidden states, $\theta_{u_{0:\mathcal{T}-1}}$ represents the sequence of actions (policy) chosen, $\theta_a$ ($=\mathbf{a}$) represents beliefs about the observation likelihood mapping, and $\theta_b$ ($=\mathbf{b}$) represents beliefs about the state transition dynamics contingent on actions. Learning corresponds to updating $\theta_a$ and $\theta_b$ across trials. 
In this framework, if the generative model and process sport the same state space dimensions, agent performance may be quantified in terms of distances between true environment dynamics ($\mathbf{A},\mathbf{B}$) and learnt environment mechanics ($\mathbf{a},\mathbf{b}$). We give more detailed accounts of this performance metric in the Annex.

\section{Simulations}

\begin{figure}
	\centering
	\includegraphics[width=\linewidth]{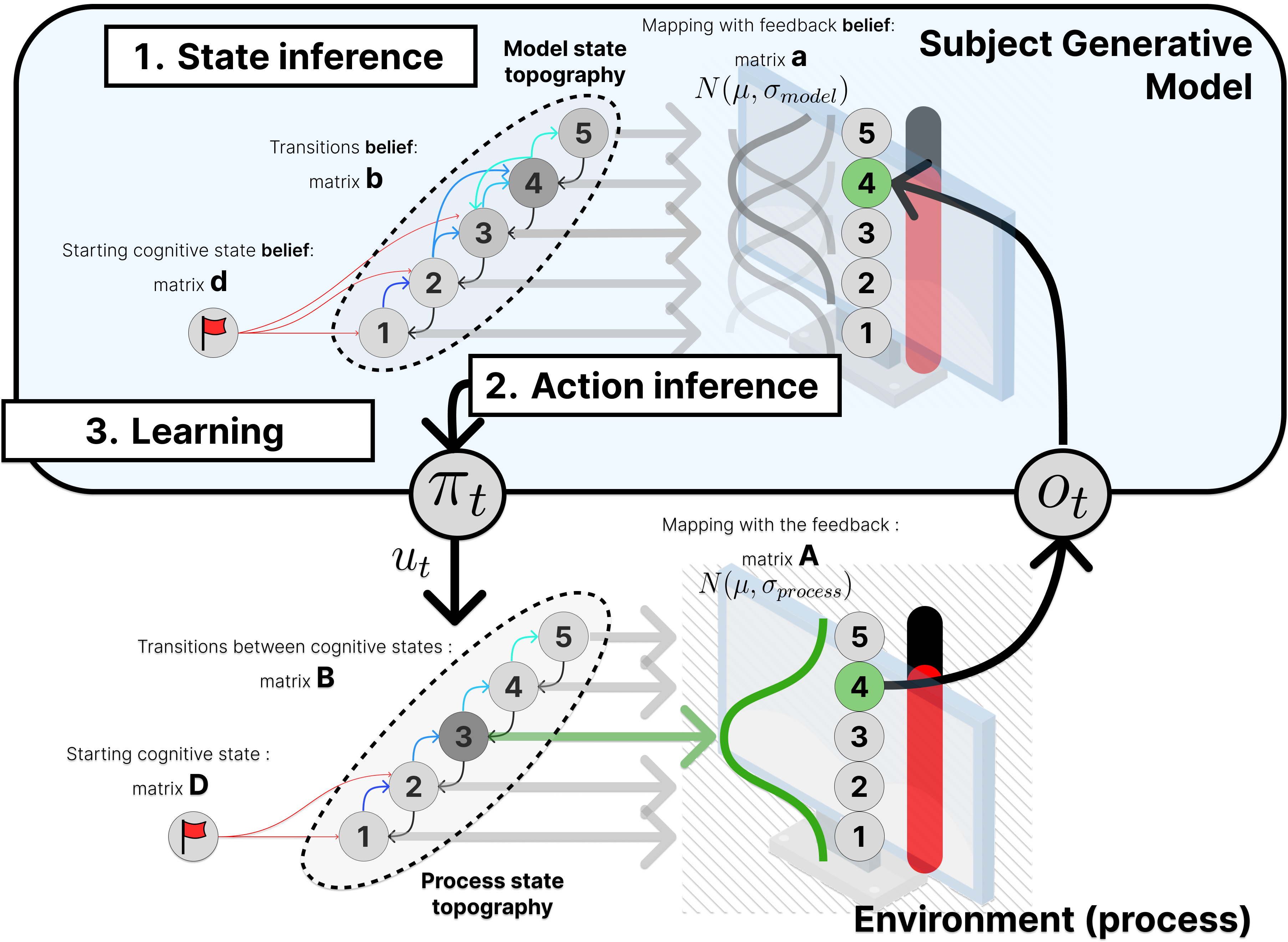}
	\caption{Full model.  The environment (generative process) forwards observations to the subject as a discrete feedback value (here $o_t=4$) and reacts to subject actions $u_t \sim \Pi_t$. In our simulations, the process state topography was a simple graph with 5 possible states and 5 possible ooutcomes. Possible state transitions are showed with blue arrows ($\mathbf{B}$) and starting state with red arrows ($\mathbf{D}$). Each state may generates outcomes based on the feedback mapping ($\mathbf{A}$) following a normal law $N(\mu,\sigma_{process})$ (unbiaised noisy feedback). Initially, the agent \textit{believes} each state is correlated with a specific feedback level following a normal law $N(\mu,\sigma_{model})$ (unbiaised noisy feedback). Given these priors, the agent attempts to learn state transitions and starting values ($\mathbf{b}$ and $\mathbf{d}$) in order to achieve higher feedback levels. This is done in 3 steps : the subject uses its priors to \textbf{1.} infer the hidden states, \textbf{2.} pick the best actions (for exploratory or exploitative purposes) and finally \textbf{3.} update its mapping from the observed succession of actions/observations.} 
	\label{fig:layout_full}
\end{figure}

\subsection{General Simulation Setup}\label{sec:sim_setup}

To explore NF dynamics using this framework, we simulated AIF agents engaged in a task requiring the regulation of a single cognitive dimension, represented by discrete latent states $\hat{s}$. The agents received feedback signal(s) $\mathbf{o}$ and possessed prior preferences $\mathbf{c}$ favoring higher feedback values. The objective was to learn a sequence of mental actions $u_t$ to transition from initial, less preferred states towards target states associated with higher feedback, potentially overcoming initially inaccurate models of perception ($\mathbf{a}$) and control ($\mathbf{b}$). Figure \ref{fig:layout_full} illustrates the general model structure. \\

\textbf{Generative Process (Environment):}
\begin{itemize}
	\item \textbf{State Space and Dynamics ($\mathbf{B}, \mathbf{D}$):} The true latent state space consisted of $N_s=5$ discrete states ($\hat{s} \in \{0, 1, 2, 3, 4\}$). Subject mental actions (e.g. performing mental imagery, inner speech, memorizing task, etc.) provoked changes in subject mental states. In all subsequent models, the agent could only move from a state $i$ to the adjacent states $i-1$ and $i+1$, or remain in the current state.
	Transitions were probabilistic: a specific "up" action could increase the state index ($\hat{s} \rightarrow \hat{s}+1$) with probability $p_{up}=0.99$ (if $\hat{s}<4$), otherwise remaining in the current state. Any other action, or the "up" action failing, resulted in remaining in the current state. Additionally, a "cognitive pull" towards a baseline state was implemented: if the agent did not successfully execute the "up" action, there was a probability $p_{rest}=0.5$ of transitioning to the next lower state ($\hat{s} \rightarrow \hat{s}-1$, if $\hat{s}>0$). These values were chosen to model a task requiring sustained effort against a baseline tendency. Agents initial state was always in the bottom 2 states, encoded by the matrix $\mathbf{D}$.
	
	\item \textbf{Observations ($Cat(\mathbf{A}) = P(o_t|\hat{s}_t)$):} The environment generated observations $o_t$ from a set of $N_o=5$ possible values ($o \in \{0, 1, 2, 3, 4\}$), based on the current (true) hidden  mental state. The primary observation modality was an \textbf{explicit feedback signal}, whose likelihood followed a discretized normal distribution centered on the true state index $\hat{s}_t$ with standard deviation $\sigma_{\textit{process}}$ : $\mathbf{A}^{\text{feedback}} \sim \text{Discretize}(N(\hat{s}_t, \sigma_{\textit{process}}))$ (we provide more information on that discretization process in the annex). The value of $\sigma_{\textit{process}}$ determined the feedback noise level and typically varied between 0 (perfect feedback) and 2.0 (very noisy feedback). A feedback noise of 5.0 could be considered as akin to a sham feedback signal (almost entirely decorrelated from regulated cognitive activity). In some simulations  (Section \ref{sec:3.4}), a second observation modality was proposed in the form of  an  \textbf{implicit (interoceptive) observation modality} ($\mathbf{A}_{\text{intero}}$). This signal also depended on the true state $\hat{s}_t$, following $\mathbf{A}^{\text{intero}} = \text{Discretize}(N(\hat{s}_t, \sigma_{\textit{process}}^{\text{intero}}))$, providing a potentially independent source of information about the latent state. The reliability of this observation source was also made to vary : $ \sigma_{\textit{process}}^{\text{intero}} \in [0.0,2.0]$.
\end{itemize} 
\textbf{Generative Model (Agent):}
\begin{itemize}
	\item \textbf{State Space and Preferences ($\mathbf{d}, \mathbf{c}$):} The agent's model assumed the same number of states ($N_s=5$) and observations ($N_o=5$). Pre-training initial state beliefs $\mathbf{d}_{\dag=0}$ were uniform. For the \textbf{transition Model ($\mathbf{b}$)}, agents started with uninformed priors about the effects of their actions. The initial transition model $\mathbf{b}_{\dag=0}$ was initialized as a matrix where each possible transition $(s_t, u_t) \rightarrow s_{t+1}$ had an equal, small probability mass, scaled by an initial concentration parameter $k_b$. Formally, the Dirichlet priors were set such that $\mathbf{b}_{\dag=0} = k_{b} \mathbb{1}$ (where $\mathbb{1}$ represents a tensor of appropriate dimensions filled with ones). $k_b$ controlled the agent's initial confidence in this uniform prior; lower values imply faster learning from experience. Throughout simulations, $k_b=0.01$ was used, reflecting low initial confidence and a prompting an exploratory drive in synthetic subjects.
	\item \textbf{Likelihood Model ($Cat(\mathbf{a}) = P(o_t|s_t)$):} The agent's model of the explicit feedback was initialized based on an assumed relationship between hidden state and provided feedback. This prior was typically $\mathbf{a}^{\text{feedback}} \sim \text{Discretize}(N(s_t, \sigma_{\textit{model}}))$. The initial confidence in this prior was controlled by a concentration parameter $k1_a$. Specifically, the initial Dirichlet counts were set via:
	\begin{equation}
		\label{eq:initial_feedback_conf}
		\mathbf{a}^{\text{feedback}}_{\dag=0} = (k1_{a} \cdot \text{Discretize}(N(s_t,\sigma_{\textit{model}})) + \epsilon \mathbb{1})
	\end{equation}
	where $\epsilon$ is a small constant (e.g., 0.01) adding minimal uniform probability to prevent zero-likelihood issues, and $k1_a$ scales the initial counts, determining resistance to updating this prior. Higher $k1_a$ means stronger initial belief.
	When the interoceptive modality was included (Section \ref{sec:3.4}), agents started with an additional completely uniform prior regarding its mapping: $\mathbf{a}^{\text{intero}}_{\dag=0} = k2_{a} \mathbb{1}$, typically with a high concentration parameter $k2_a = 10.0$, reflecting high initial confidence about the uncertain meaning of this signal relative to the cognitive state. This reflects an observation modality to which we're accustomed but which mapping with the targeted cognitive state has not been noticed until the training.
	\begin{equation}
		\label{eq:initial_io_prior_formal} 
		\mathbf{a}^{\text{intero}}_{\dag=0} = k2_{a} \cdot \mathbb{1}
	\end{equation}
	\item \textbf{Preferences and habits} : Preferences $\mathbf{c}$ were set to favor higher feedback values linearly (highest preference for $o=4$, lowest for $o=0$).  Habits $\mathbf{e}$ were initially uniform and were not learnt in these simulations.
\end{itemize}

Note that for simplicity's sake and to allow for the direct comparison of true and learnt environment rules,this study's generative processes and models will always feature similar state space sizes, though a mismatch between regulated state space and subject representation is an interesting modelling prospect (see \ref{sec:discuss}).\\

Key parameters manipulated across simulations are summarized in Table \ref{tab:simulated_index}. Unless otherwise specified, agents learn both transition ($\mathbf{b}$) and likelihood ($\mathbf{a}$) models over trials. Due to the stochastic nature of the processes, N = 10 agent instances were simulated for each parameter set to ensure robustness of conclusions.

\begin{table}[!ht]
	\begin{center}        
		\begin{tabular}{||m{10em} | m{25em} | m{10em}||}
			\hline
			\textbf{Model parameters} & \textbf{Parameters} & \textbf{Used in simulations} \\
			\hline \hline
			\textbf{M1} : Topography of the hidden states of the generative model / process  &  One cognitive dimension. To get from a state to the next, the agent must pick an action depending on its starting state. & M1 was used in all simulations, but alternative topographies are described in annex. \\
			\hline
			Variance of the feedback generative process $\sigma_{process}$ & Set the hidden rule behind the generation of a specific feedback. How noisy compared to the true cognitive state ? & All. \\
			\hline
			Variance of the interoceptive generative process $\sigma_{process}^{intero}$ & Set the hidden rule behind the generation of a specific interoceptive signal. How noisy compared to the true cognitive state ? & \ref{sec:3.4}.\\
			\hline
			Variance of the subject generative model of the feedback $\sigma_{model}$ & Set the expectations of the subject regarding the reliability of feedback when the training starts. & \ref{sec:3.2},\ref{sec:3.3} and \ref{sec:3.4}. \\
			\hline
			Initial confidence of the subject in its action model $k1_b$ & How much weight the agent gives to its initial action model. The subject starts with a flat / noisy prior regarding its mental actions. We set how confident the agent is regarding this mapping with this factor. Higher values mean the subject is much less prone to changing its transition beliefs and may require a longer training to achieve results.& $k1_b=0.01$ in all simulations .\\
			\hline
			Confidence of the subject in its feedback model $k1_a$ / in its interoceptive mapping $k2_a$ & How much weight the agent gives to its initial external feedback model / internal observation modality. Higher values mean the subject gives more credit to those prior instructions / beliefs and is much less prone to changing its observation model despite contradicting evidence. \textit{These parameter only had an influence when the simulated agents learned observation mappings.} & \ref{sec:3.3} and \ref{sec:3.4} [feedback] \ref{sec:3.4}[interoception]. \\
			\hline
			Presence of an interoceptive observation modality (\textbf{IO}) & If agents were given an alternative observation signal to complement the external (BCI based) feedback. & \ref{sec:3.4}. \\
			\hline
		\end{tabular}
	\end{center}
	\caption{A general sum-up of the parameters we leverage in our simulations.}
	\label{tab:simulated_index}
\end{table}

\subsection{Tackling BCI questions with simulations}

The AIF framework enables simulating diverse NF scenarios to address specific questions about training mechanisms. We focused on the following key aspects:

\subsubsection{How does the noise of the feedback affect the training ?}\label{sec:3.1}

To isolate the impact of external feedback quality, we first simulated agents whose perception model $\mathbf{a}$ was fixed throughout training and assumed perfect feedback fidelity. Specifically, agents operated with $\mathbf{a} = \mathbb{I}_{N_o}$, meaning they fully trusted the feedback as a direct readout of their state ($s_t = o_t$). Learning was restricted to the action model $\mathbf{b}$, simplifying the inference problem (Eq. \ref{eq:nf_inf_pb}) to updating only $\theta_b$:

\begin{equation}
    \label{eq:nf_inf_pb_act}
    \tag{Partial NF Inference problem} 
    \begin{aligned}
        \theta_{NF}^{\dag} &= (\theta_s, \theta_u ,\theta_{\alpha}) \\
        &= (\theta_{s},\theta_{u},\cancel{\theta_{a}},\underbrace{\theta_{b}}_{\text{ACTION belief}})
    \end{aligned}
\end{equation}

Of course, that simplification did not imply anything concerning the true feedback matrix of the generative process $\mathbf{A}$. We varied the true noise level of the generative process feedback, $\sigma_{process}$, across conditions representing:
\begin{itemize}
    \item \textbf{High-fidelity feedback} ($\sigma_{process} \approx 0.1$): True mapping $\mathbf{A} \approx \mathbb{I}_{N_s}$. Agent's assumption holds true.
    \item\textbf{Noisy feedback} ($\sigma_{process} = 0.5$): $\mathbf{A}$ is informative but imperfect, representing typical BCI variability.
    \item \textbf{Sham feedback} ($\sigma_{process} = 5.0$): $\mathbf{A}$ is nearly uniform, providing minimal information about $\hat{s}_t$.
\end{itemize}
Training trajectories (states, feedback, learned $\mathbf{b}$) were analyzed to assess how objective feedback noise affects learning when the agent assumes perfect feedback.

\subsubsection{Influence of Initial Expectations about Feedback Quality}\label{sec:3.2}

In a second time, we questionned how the agent's initial belief about feedback reliability influenced training, while still keeping the perception model $\mathbf{a}$ fixed (no learning of $\theta_a$). Unlike Section \ref{sec:3.1}, the fixed $\mathbf{a}$ was not necessarily identity but reflected varying levels of assumed noise, parameterized by $\sigma_{model}$ in the initial setup $\mathbf{a} \sim \text{Discretize}(N(s_t, \sigma_{model}))$. We varied both the true feedback noise $\sigma_{process}$ (as before) and the agent's fixed assumption $\sigma_{model}$ (from near-perfect $\approx 0.01$ to very unreliable $= 2.5$) to understand how congruence between reality and expectation affects mental action learning ($\mathbf{b}$).

\subsubsection{Modelling changing subject confidence in the feedback}\label{sec:3.3}

We then enabled agents to learn their perception model $\mathbf{a}$ alongside their action model $\mathbf{b}$, addressing the full inference problem (Eq. \ref{eq:nf_inf_pb}). This simulation explored how agents adapt their beliefs about feedback reliability based on experience. We examined the interplay between: a) the true feedback noise ($\sigma_{process}$), b) the agent's initial belief about this noise ($\sigma_{model}$ used in Eq. \ref{eq:initial_feedback_conf}), and c) the agent's initial confidence in this belief, set by the concentration parameter $k1_a$ (varied from 1.0 to 10.0). This setup models situations where subjects might initially trust or distrust experimenter instructions about feedback meaning and subsequently revise these beliefs. 

Training outcomes and the evolution of the learned perception model $\mathbf{a}$ were assessed. We ran similar simulations as in section \ref{sec:3.2} with the true feedback noise $\sigma_{process}$ and the initial subject feedback model $\sigma_{model}$ both varying between 0.01 and 2.0. and plotted the evolution of the cognitive states / models of the subjects in order to investigate how initial beliefs ($\sigma_{model}$) and confidence ($k1_a$) interact with true feedback noise ($\sigma_{process}$) when agents can adapt their perception model.

\subsubsection{Simulating training with Interoceptive Learning}\label{sec:3.4}

Finally, to investigate potential mechanisms for NF skill generalization (i.e., self-regulation without external feedback), we introduced a second, implicit observation modality intended to represent interoception (\textbf{IO} agents). These agents received both the standard external feedback (characterized by $\sigma_{process}$, initial model $\sigma_{model}$, confidence $k1_a$) and an internal signal (characterized by $\sigma_{process}^{\text{intero}}$ and confidence $k2_a$). In these simulations, we compared three typical interoceptive signals : highly informative ($\sigma_{process}^{\text{intero}}=0.0$), highly uninformative ($\sigma_{process}^{\text{intero}} = 2.0$) and intermediate ($\sigma_{process}^{\text{intero}} = 0.9$).
Importantly, here the subject was much more accustomed to the latter source of information (translating into a high $k2_a$ initial confidence parameter set to 10.0 in our simulations) but had not related these internal observations to their current cognitive states (flat initial prior):  agents started with highly uncertain priors about the interoceptive mapping ($\mathbf{a}_{\dag=0}^{\text{intero}}$ uniform, Eq. \ref{eq:initial_io_prior_formal})
They were able to learn both $\mathbf{a}_{\text{feedback}}$ and $\mathbf{a}_{\text{intero}}$ alongside $\mathbf{b}$. We compared the performance and learned models of IO agents versus agents receiving only external feedback, across various levels of external feedback noise ($\sigma_{process}$) and initial beliefs ($\sigma_{model}$). The accuracy of the learned interoceptive mapping $\mathbf{a}_{\text{intero}}$ at the end of training was used as a proxy for the potential for post-training generalization.

\section{Results}\label{section:results}
 
The figures presented in this section and the simulations used to generate the results are available at \url{https://github.com/Erresthor/ActivPynference_Public/tree/main/paper_scripts/paper_ActiveInference_BCI}

\subsection{How does the noise of the feedback affect the training ?}\label{sec:results_3.2.1}

\begin{figure*}
    \centering
    \begin{subfigure}[b]{0.95\textwidth}
        \centering
        \includegraphics[width=\linewidth]{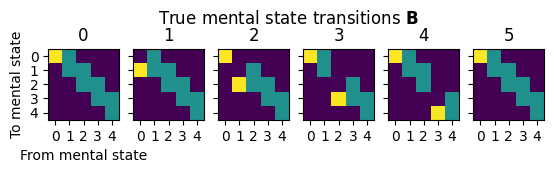}
        \caption{True effect of mental actions.}\label{fig:agent_perfect_true_matrices}
    \end{subfigure}
    \hfill
    \begin{subfigure}[b]{0.475\textwidth}
        \centering
        \includegraphics[width=\linewidth]{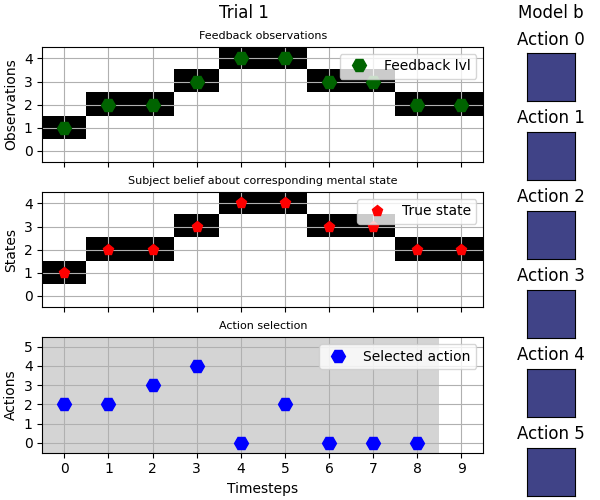}
        \caption{Trial 1.}\label{fig:agent_perfect_1}
    \end{subfigure}
    \hfill
    \begin{subfigure}[b]{0.475\textwidth}  
        \centering 
        \includegraphics[width=\linewidth]{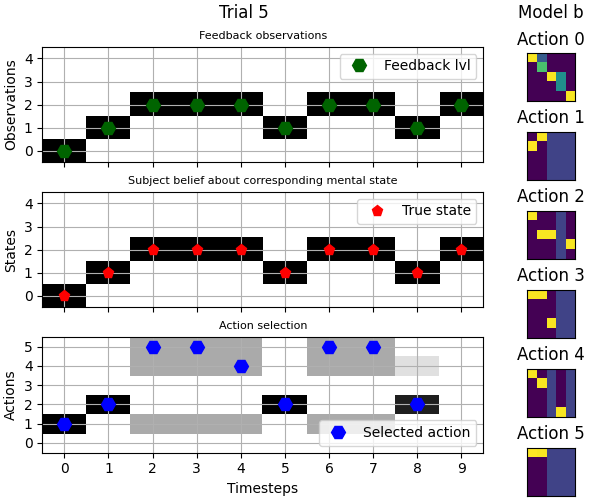}
        \caption{Trial 5.}\label{fig:agent_perfect_2}
    \end{subfigure}
    \vskip\baselineskip
    \begin{subfigure}[b]{0.475\textwidth}   
        \centering 
        \includegraphics[width=\linewidth]{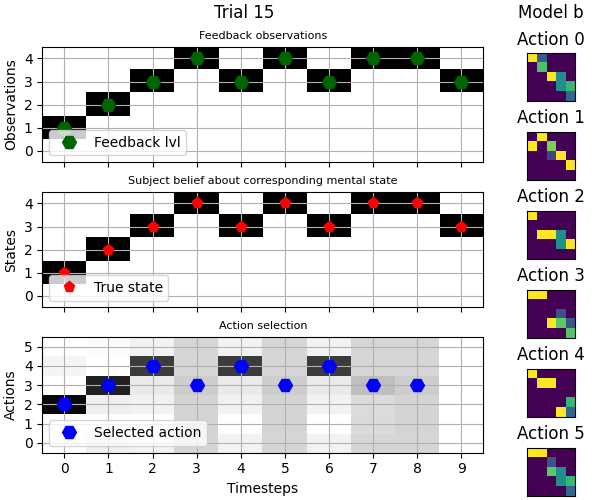}
        \caption{Trial 15.}\label{fig:agent_perfect_3}
    \end{subfigure}
    \hfill
    \begin{subfigure}[b]{0.475\textwidth}   
        \centering 
        \includegraphics[width=\linewidth]{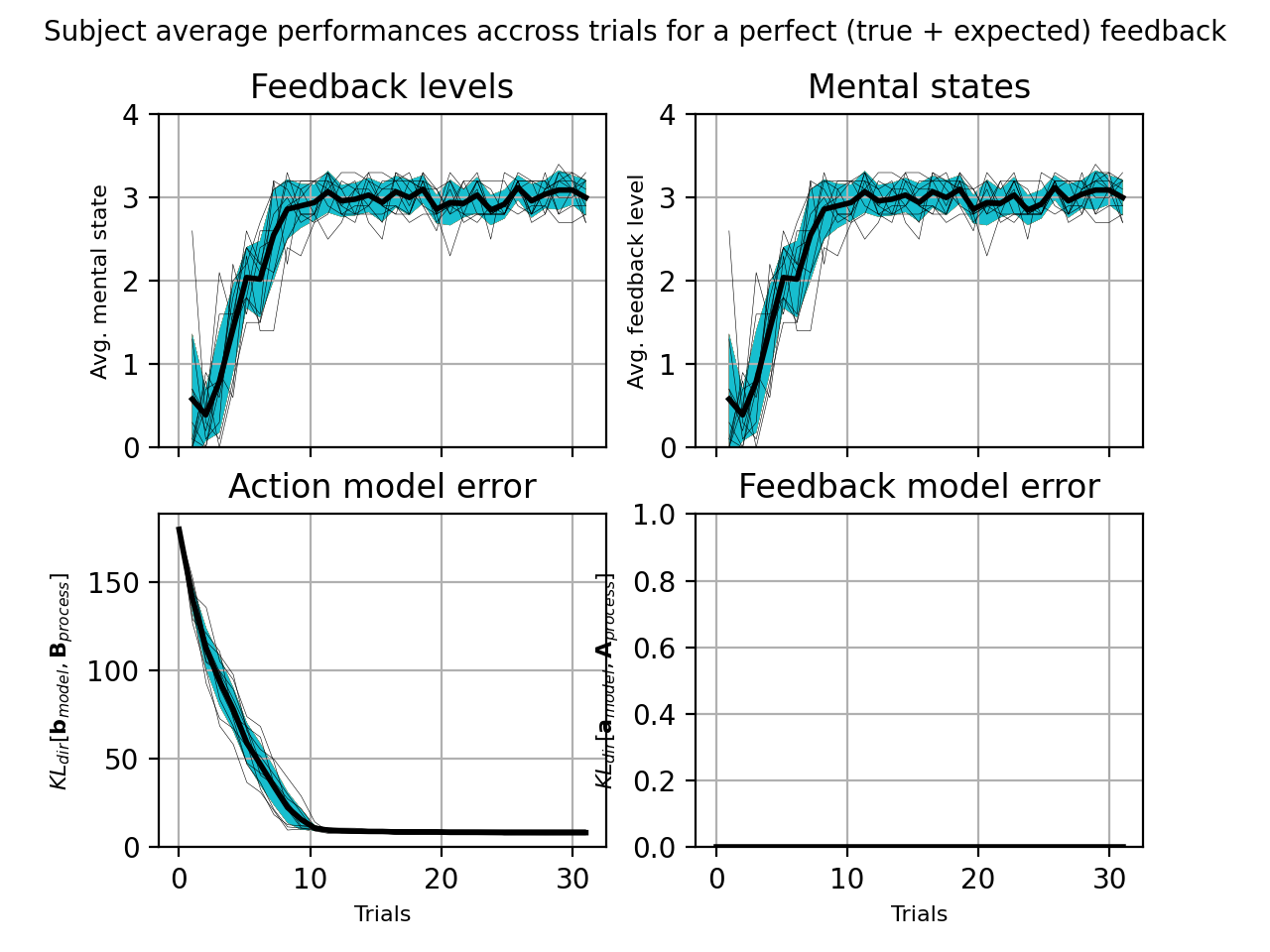}
        \caption{Training curves.}\label{fig:agent_perfect_4}
    \end{subfigure}
    \caption{Simulated trials of 10 agents with a perfect model of a perfect feedback ($\mathbf{a} = \mathbf{A} = \mathbb{I}_5$) but no initial model of mental actions. The true effect of those mental actions is shown in \ref{fig:agent_perfect_true_matrices} (each action has a different effect depending on the state of the agent (x-axis), and leads the subject to another state (y-axis). We represent artificial agent behaviour and learning across trials. We show how individual trials are conducted (\ref{fig:agent_perfect_1}, \ref{fig:agent_perfect_2}, \ref{fig:agent_perfect_3}), with the observations (first row), subject mental states (second row) and action taken (third row). The colored dots represent the true (observed/hidden/selected) values and the black-and-white background represents the distributions from which they were sampled (top/bottom row) and the subject infered current cognitive state (middle row). We also plot the subject transition model during the trial to explain why it may have picked some actions rather than others. Because this representation doesn't easily qualify several subjects accross 100+ trials, we also displayed training curves of the subjects (\ref{fig:agent_perfect_4}). It consists in the evolution of their average feedback level, mental state, as well as transition and feedback model error accross trials. The subjects were initially very uncertain about which action to take in the initial trials, but their perception was always certain and the subjects knew exactly what were their cognitive state at any point. Because the action model of the subject started off too incomplete, the agent first engaged in explorative behaviour to gather informations and improve its model. It quickly made sense of the available data and built a good model of its mental actions. By trial 10, the model of the agent is good enough that it only focuses on achieving high feedback levels.} 
    \label{fig:agent_perfect}
\end{figure*}

\begin{figure*}
    \centering
    \begin{subfigure}[b]{0.475\textwidth}
        \centering
        \includegraphics[width=\linewidth]{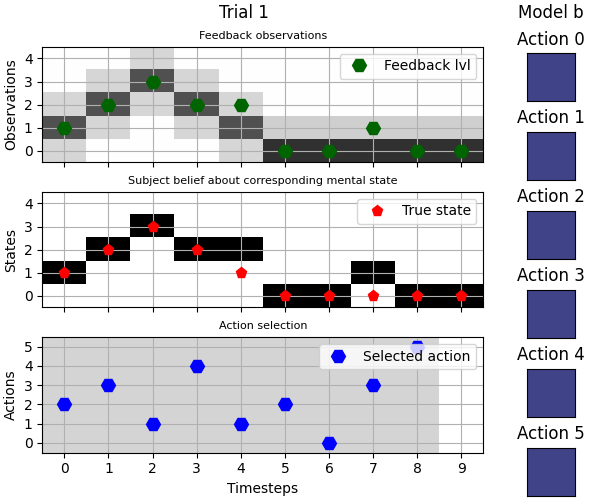}
        \caption{Trial 1.}\label{fig:agent_noisy_1}
    \end{subfigure}
    \hfill
    \begin{subfigure}[b]{0.475\textwidth}  
        \centering 
        \includegraphics[width=\linewidth]{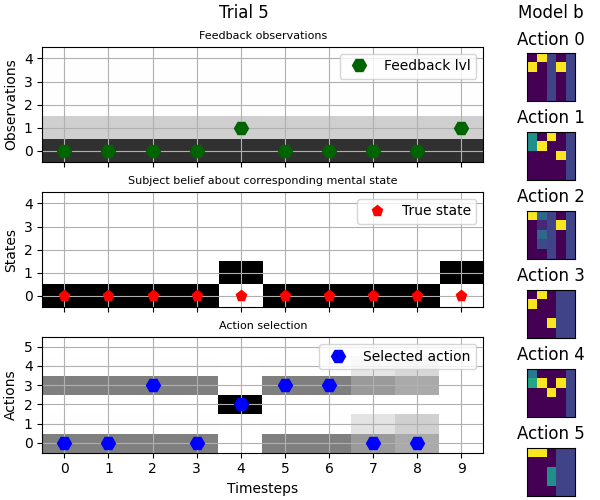}
        \caption{Trial 5.}\label{fig:agent_noisy_2}
    \end{subfigure}
    \vskip\baselineskip
    \begin{subfigure}[b]{0.475\textwidth}   
        \centering 
        \includegraphics[width=\linewidth]{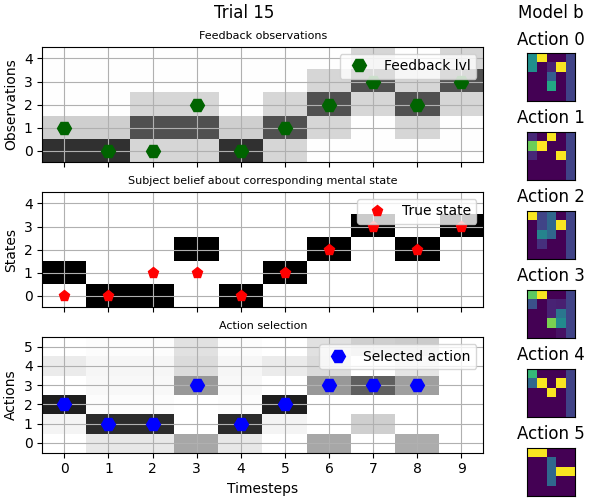}
        \caption{Trial 15.}\label{fig:agent_noisy_3}
    \end{subfigure}
    \hfill
    \begin{subfigure}[b]{0.475\textwidth}   
        \centering 
        \includegraphics[width=\linewidth]{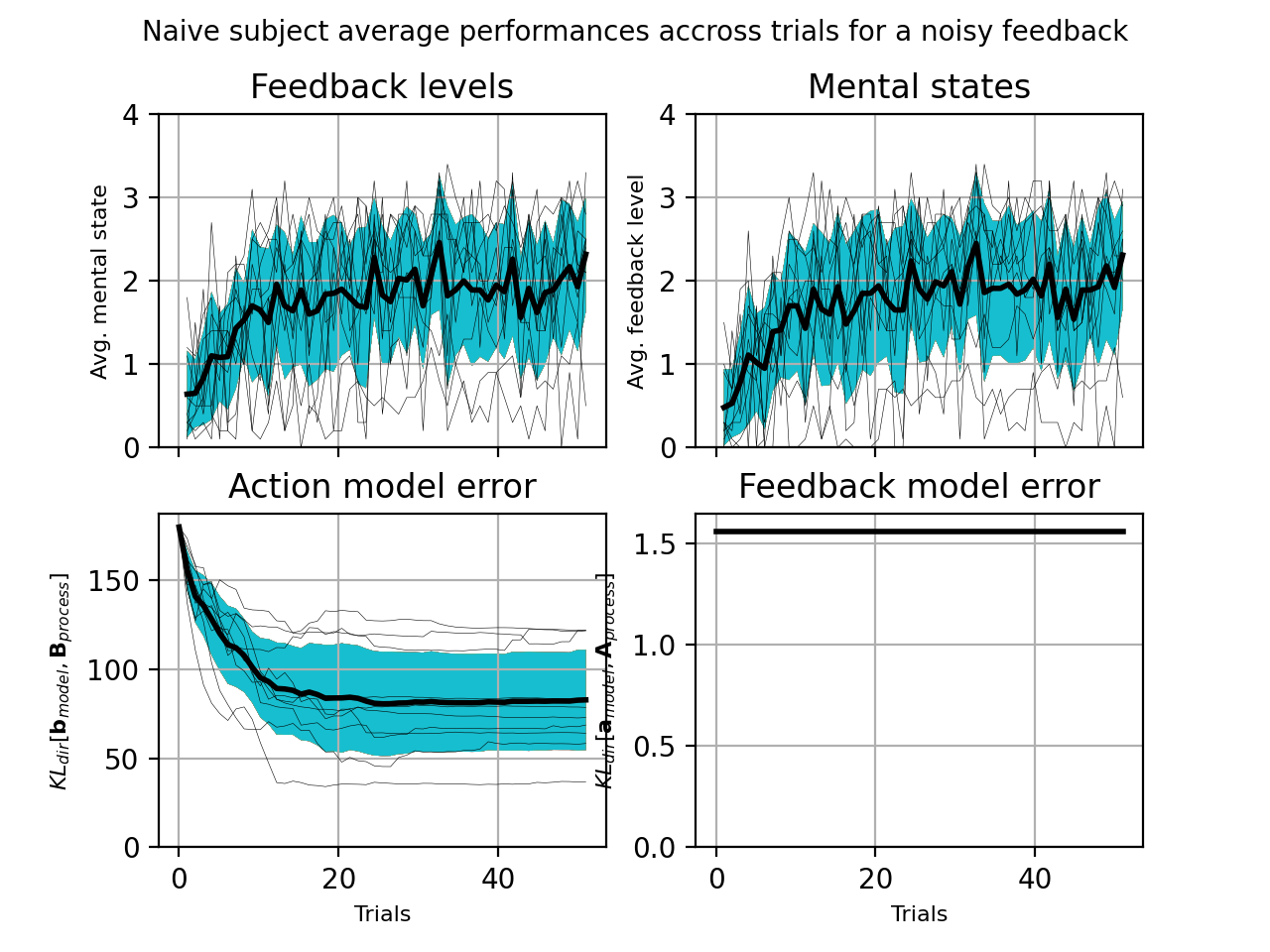}
        \caption{Training curves.}\label{fig:agent_noisy_4}
    \end{subfigure}
    \caption{Simulated trials of 10 agents with a naive model ($\sigma_{model}=0.01$) and a noisy feedback ($\sigma_{model}=0.5$). In contrast to figure \ref{fig:agent_perfect}, observations were sampled from a noiser distribution and created mismatches between subject inferences and true mental states (\ref{fig:agent_noisy_2}), thus leading to lower learnt transition model quality and regulation performances.} 
    \label{fig:agent_noisy_biomarker}
\end{figure*}

\begin{figure*}
    \centering
    \includegraphics[width=\linewidth]{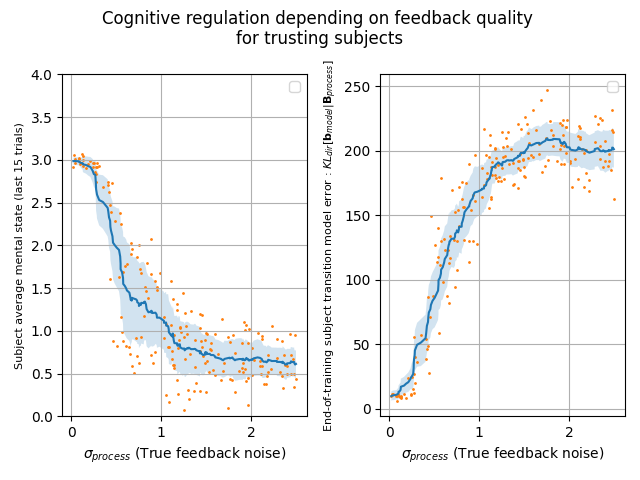} 
    \caption{Subject performance metrics at the end of the training for various levels of biomarker noise. We plot the average mental state achieved during the last 15 trials of the (100 trials) training [left] as well as the error of the transition model learnt by the subjects at the end of the training [right]. Each point corresponds to the a single agent which believed that the feedback was perfect ($\mathbf{a}=\mathbb{I}_5$) but was provided with a true feedback noise$\sigma_{process} \in [0.0,2.5]$ (x-axis).  The noise associated with the chosen biomarker has a huge influence on agent performance and training efficacy, and subject ability to learn from the feedback fall off dramatically after a certain threshold ($\simeq 0.3$)}
    \label{fig:biomarker_noise_plot}
\end{figure*}

\textbf{High-fidelity feedback :} We first simulated a set of 10 artificial agents using a perfect feedback to self-regulate. Figures \ref{fig:agent_perfect_1},\ref{fig:agent_perfect_2},\ref{fig:agent_perfect_3} show a representation of one of these agents internal variables during three trials (1, 5 and 15). The figures upper rows show observations (observed outcome and sampled distribution), the middle rows show the hidden states (true hidden value and subject inference) and the lower rows  depict agent posterior over actions (how desirable a given action appears to the subject) and selected action (in this case the action among the biggest action posteriors). In all three trials, the agents were embedded with a perfect perception model, thus making perception a trivial task and explaining the certainty with which the agents inferred their hidden mental states given feedback. The subject used the initial trials to test actions and learn the transition rules. By trial 15 however, the action model was informed enough to prompt exploitative behaviour in the subject (always taking the shortest path towards the wanted state). A clear explorative/exploitative shift is noticeable at the trial level. Figure \ref{fig:agent_perfect_4} gives an account of the same situation, but at the whole training scale. This time, the results of all 10 subjects are shown and subject performances are quantified using a set of metrics (defined in Annex) to allow for better readability. We plot the average feedback level and cognitive state achieved by the agents during the last 15 trials of the training. We also plot the summed KL-divergence between their transition and observation models, and their true counterparts. Again we notice the explorative/exploitative shift around trial 10.

\textbf{Unbiased but noisy feedback :} Figure \ref{fig:agent_noisy_biomarker} shows how agent performances are affected by a noisy feedback with no bias. This may describe various parts of the BCI pipeline, and especially the biomarker used. To question the usefulness of noisy biomarkers, a noisier feedback ($\sigma{process} = 0.5$) was provided to the subjects. As previously expected, higher levels of noise led to lower overall feedback and cognitive levels and a final action model of poorer quality. Despite a rather low amount of noise, performances at the end of the training were significantly affected and it prompted us to simulate a wider range of biomarker noises. 

\textbf{Sham feedback :} The agents who attempted to learn to regulate their hidden states using sham feedback did not manage to improve their performances significantly. This comes as no surprise as we did not implement any non specific mechanism that would explain successful regulation. The associated figure is available in Annex.\\

Figure \ref{fig:biomarker_noise_plot}  shows the training efficacy fall-off depending on the noise of the selected biomarker.  Basing ourselves on the previous simulations, we plot the training of subjects with feedbacks of varying qualities. We selected the true feedback noise randomly between $\sigma_{process} = 0.01$ (perfect feedback) and $\sigma_{process} = 2.5$ (sham feedback). For all data points, the subject expected the feedback to be perfect ($\sigma_{model} = 0.01$) and did not update this belief during training. We plot the average cognitive regulation achieved by the artificial subjects at the end of the training (avg. last 20 trials) and the quality of their model. As expected, a lower biomarker quality leads to worse self-regulation performances. The efficacy of the training falls of dramatically for $\sigma_{process} \simeq 0.5$, eventually rendering the whole training paradigm nearly useless around $\sigma_{process} \simeq 1.5$. This shows that when navigating a complex environment with sparse feedback, the reliability of the signal is primordial to achieve self-regulation. In the next section, we explore if this conclusion can be somewhat mitigated by agent explicit modelling of the noise.

\subsection{Influence of Initial Expectations about Feedback Quality}

\begin{figure*}
    \centering
    \begin{subfigure}[b]{0.475\textwidth}
        \centering
        \includegraphics[width=\linewidth]{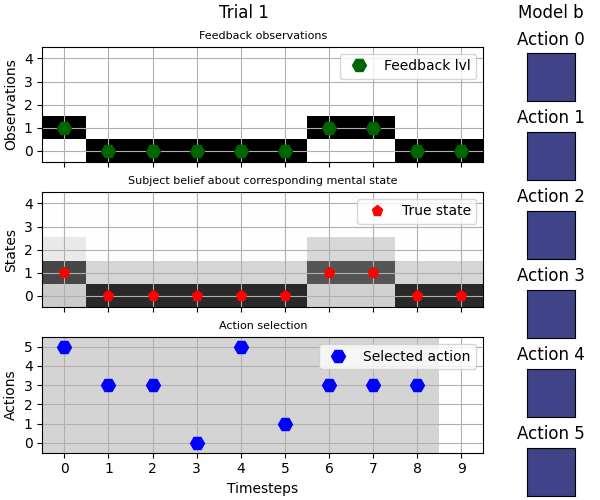}
        \caption{Trial 1.}\label{fig:agent_cautious_1}
    \end{subfigure}
    \hfill
    \begin{subfigure}[b]{0.475\textwidth}  
        \centering 
        \includegraphics[width=\linewidth]{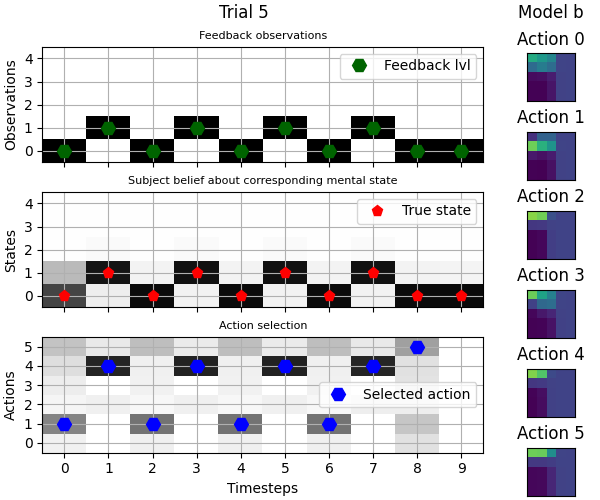}
        \caption{Trial 5.}\label{fig:agent_cautious_2}
    \end{subfigure}
    \vskip\baselineskip
    \begin{subfigure}[b]{0.475\textwidth}   
        \centering 
        \includegraphics[width=\linewidth]{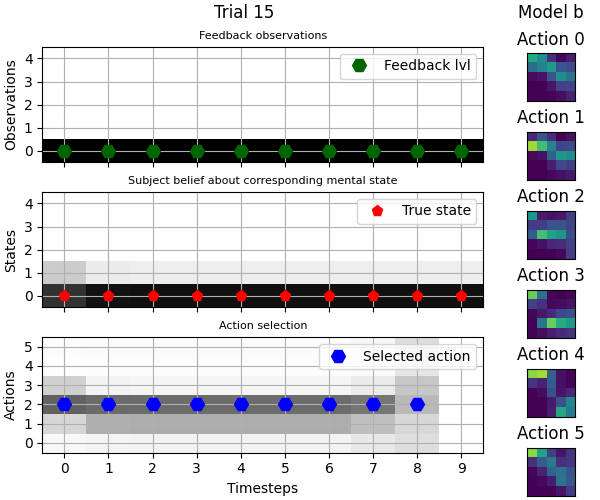}
        \caption{Trial 15.}\label{fig:agent_cautious_3}
    \end{subfigure}
    \hfill
    \begin{subfigure}[b]{0.475\textwidth}   
        \centering 
        \includegraphics[width=\linewidth]{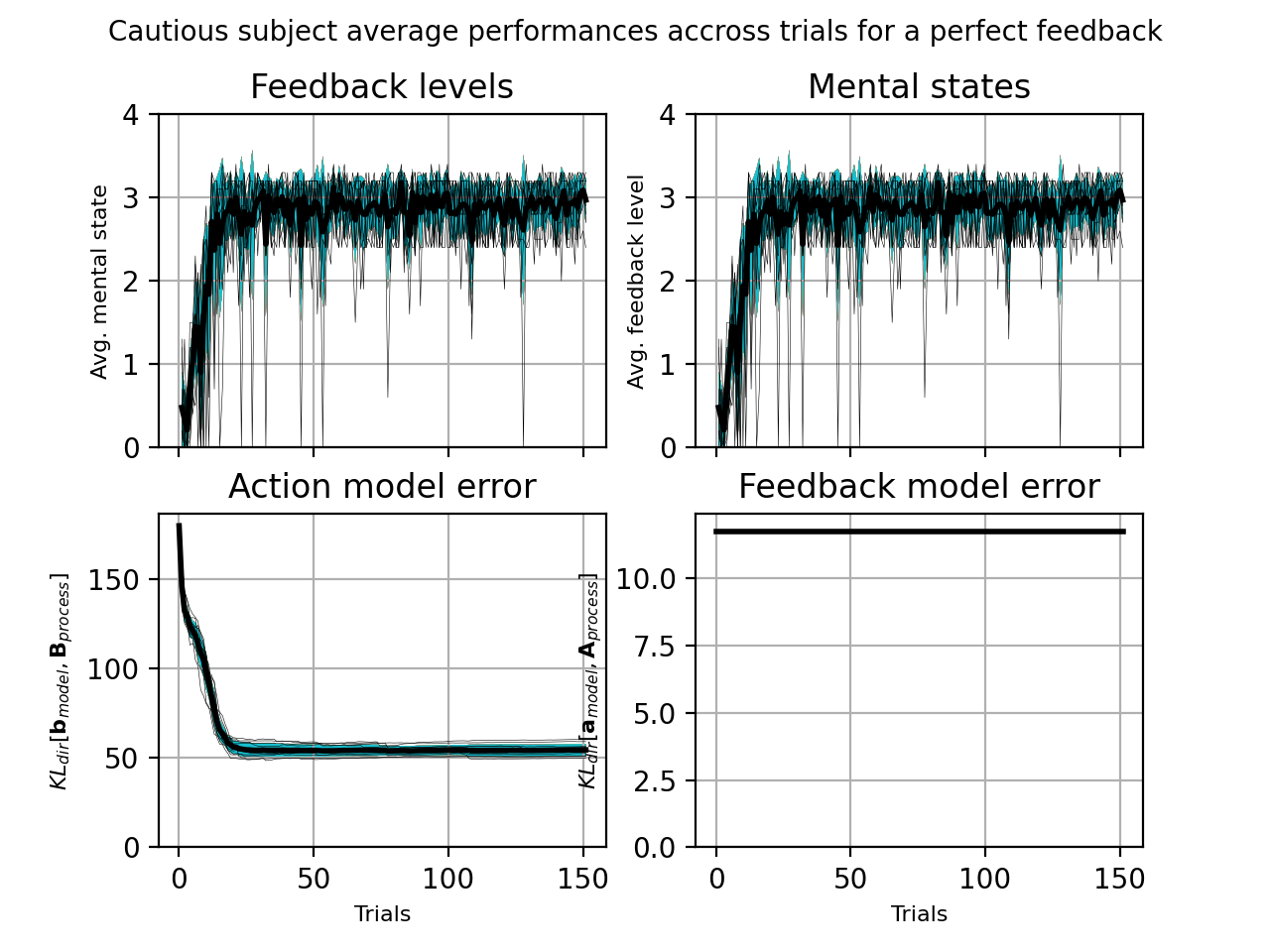}
        \caption{Training curves.}\label{fig:agent_cautious_4}
    \end{subfigure}
    \caption{Simulated trials of 10 agents with a cautious model ($\sigma_{model}=0.5$) and a perfect feedback ($\sigma_{model}=0.01$). Here, subjects explicitly doubted the meaning of a specific feedback value, assuming that it could have been generated by a range of cognitive states (driven by $\sigma_{model}$). Although the subjects were overly cautious about the feedback, they still managed to self-regulate optimally.} 
    \label{fig:agent_cautious}
\end{figure*}

\begin{figure*}
	\centering
	\begin{subfigure}[b]{0.475\textwidth}
		\centering
		\includegraphics[width=\linewidth]{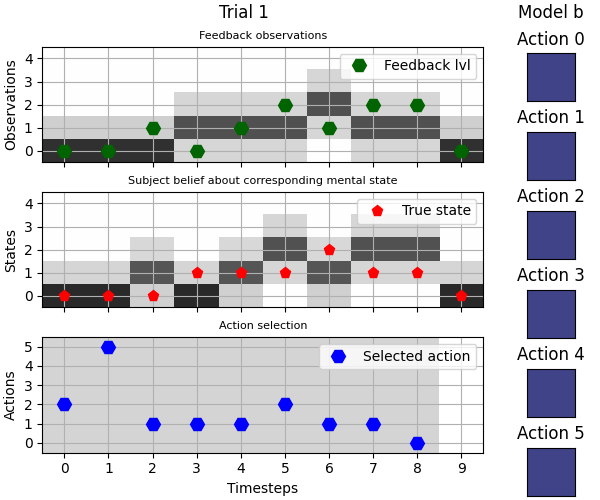}
		\caption{Trial 1.}\label{fig:agent_cautious2_1}
	\end{subfigure}
	\hfill
	\begin{subfigure}[b]{0.475\textwidth}  
		\centering 
		\includegraphics[width=\linewidth]{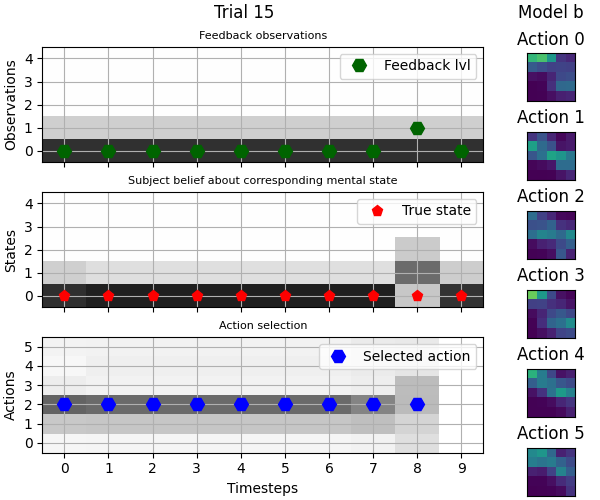}
		\caption{Trial 15.}\label{fig:agent_cautious2_2}
	\end{subfigure}
	\vskip\baselineskip
	\begin{subfigure}[b]{0.475\textwidth}   
		\centering 
		\includegraphics[width=\linewidth]{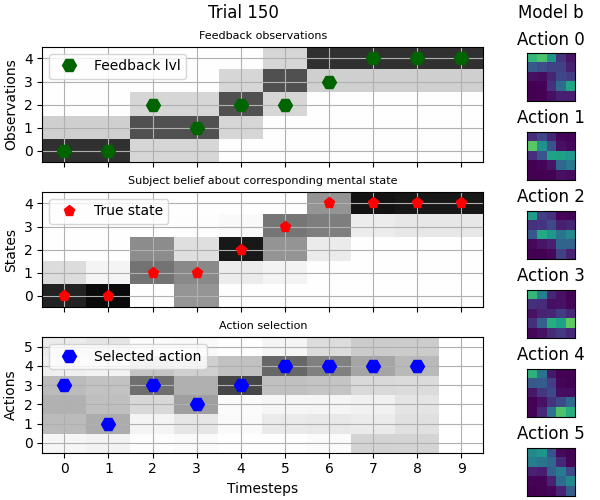}
		\caption{Trial 150.}\label{fig:agent_cautious2_3}
	\end{subfigure}
	\hfill
	\begin{subfigure}[b]{0.475\textwidth}   
		\centering 
		\includegraphics[width=\linewidth]{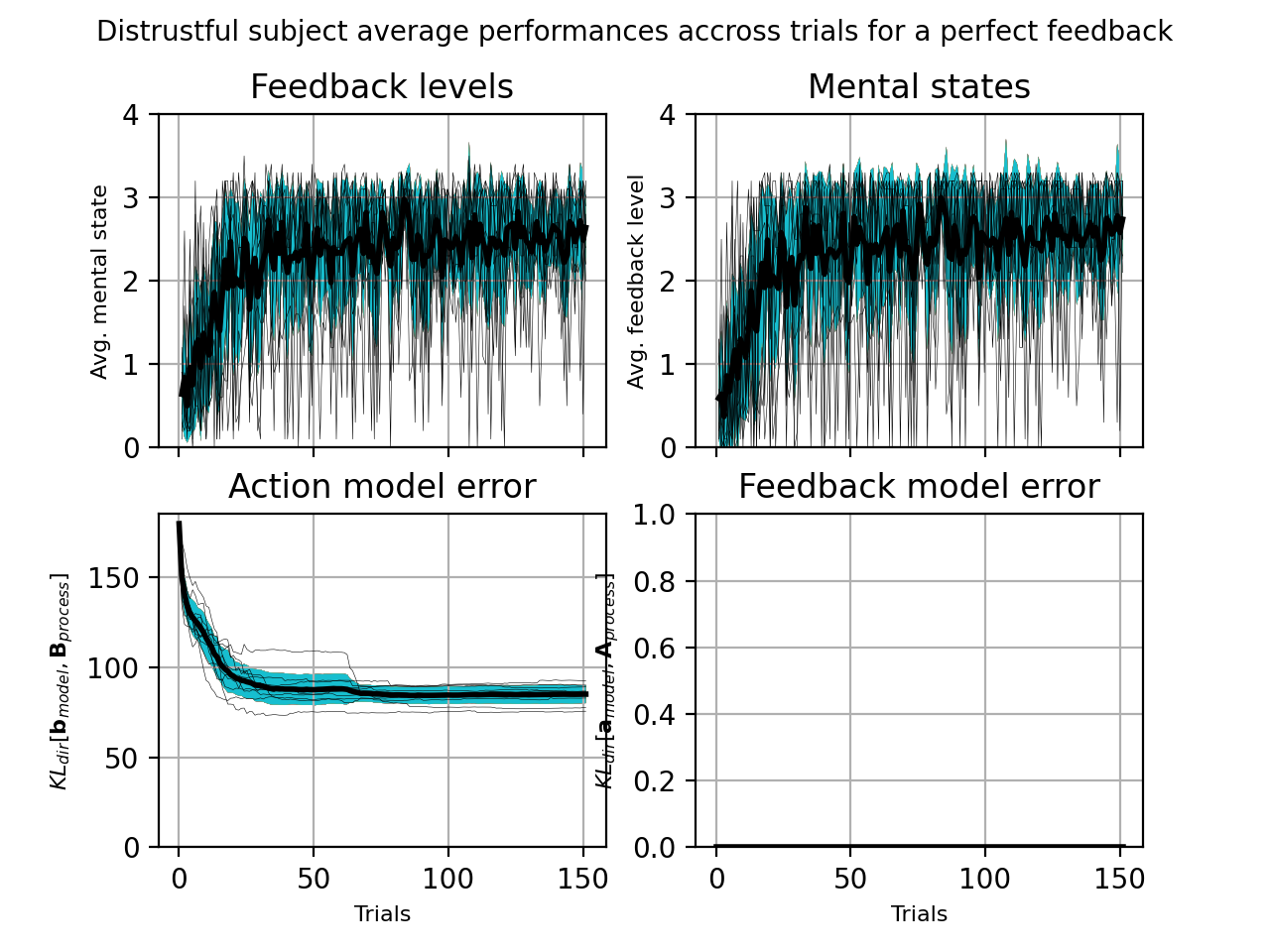}
		\caption{Training curves.}\label{fig:agent_cautious2_4}
	\end{subfigure}
	\caption{Simulated trials of 10 agents with a cautious model ($\sigma_{model}=0.5$) and a noisy feedback ($\sigma_{model}=0.5$). Here, subjects explicitly doubted the meaning of a specific feedback value, assuming that it could have been generated by a range of cognitive states (driven by $\sigma_{model}$). The explicit modelling of the noise inherent to the system allowed subject to eventually self-regulate much better than when they believed the feedback was perfect (\ref{fig:agent_noisy_biomarker}), underlining the importance of subject priors (and experimenter instructions) in a noisy context.}
	\label{fig:agent_cautious_noisy}
\end{figure*}

\begin{figure}
    \centering
    \includegraphics[width=\linewidth]{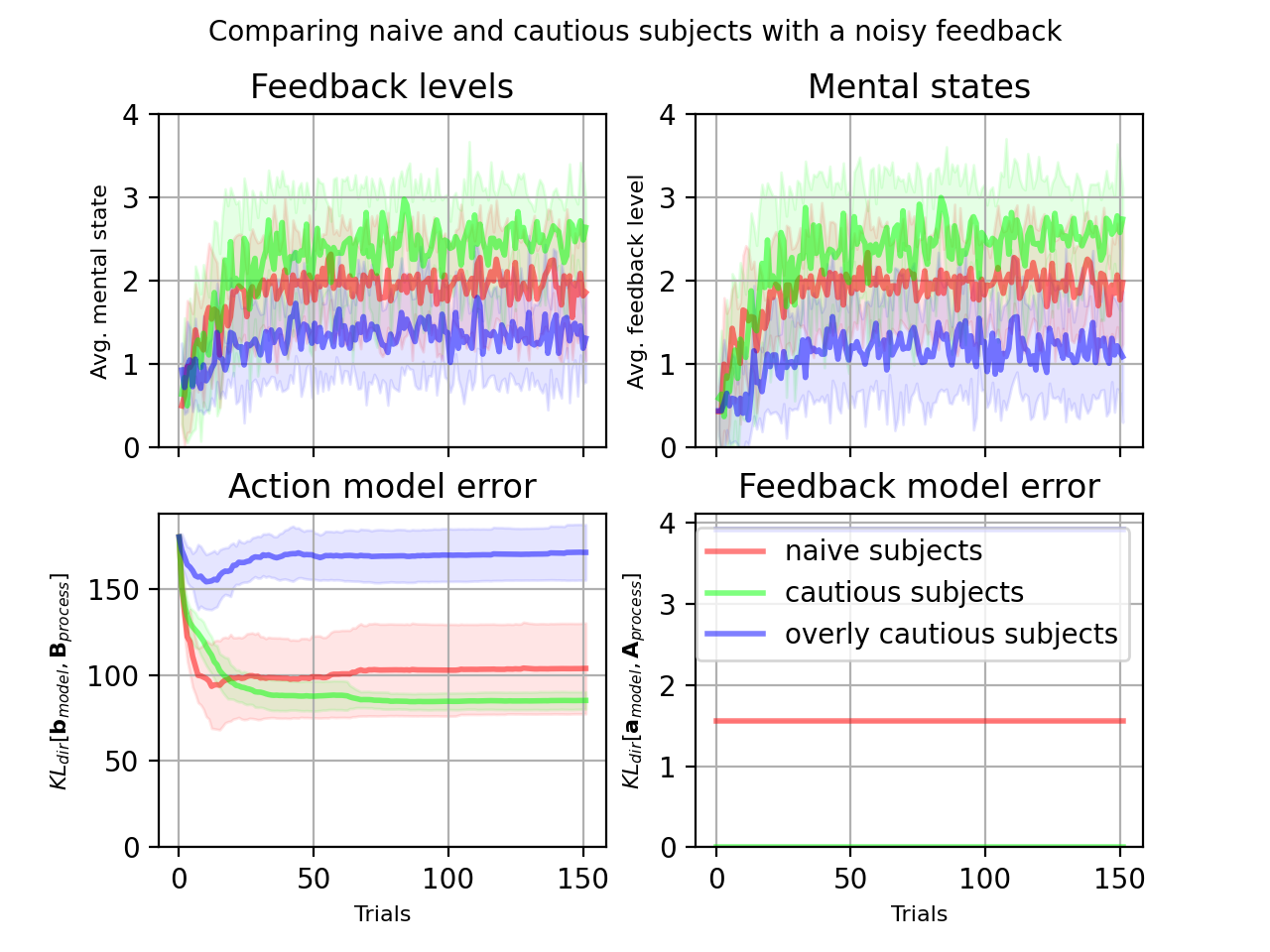}
    \caption{Compared training curves of two groups of 10 agents provided with a noisy feedback ($\sigma_{process}=0.5$). The "naive" group (blue) expected the feedback to be perfect ($\sigma_{model}=0.01$), whereas the "cautious" group had less optimistic expectations  ($\sigma_{model}=0.5$), much closer to the truth. Due to being less sensible to noisy observations, the cautious group was less prone to model overfitting and eventually, albeit less quickly, developed a better understanding of their mental actions. } \label{fig:compared_training_curve}
\end{figure}

\begin{figure*}
    \centering
    \begin{subfigure}[b]{\textwidth}
        \centering
        \includegraphics[width=\linewidth]{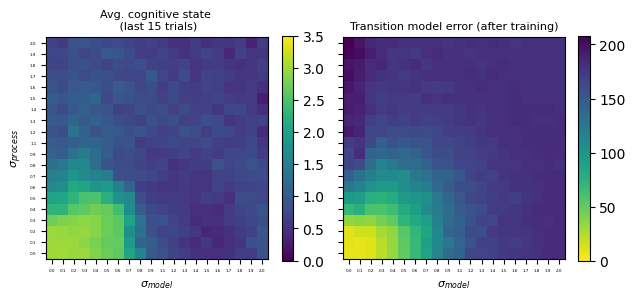}
        \caption{Agent performance map for a given expected feedback noise (x-axis) and true feedback noise (y-axis).}\label{fig:agent_double_var_nolearn_1}
    \end{subfigure}
    \vskip\baselineskip
    \begin{subfigure}[b]{\textwidth}  
        \centering 
        \includegraphics[width=\linewidth]{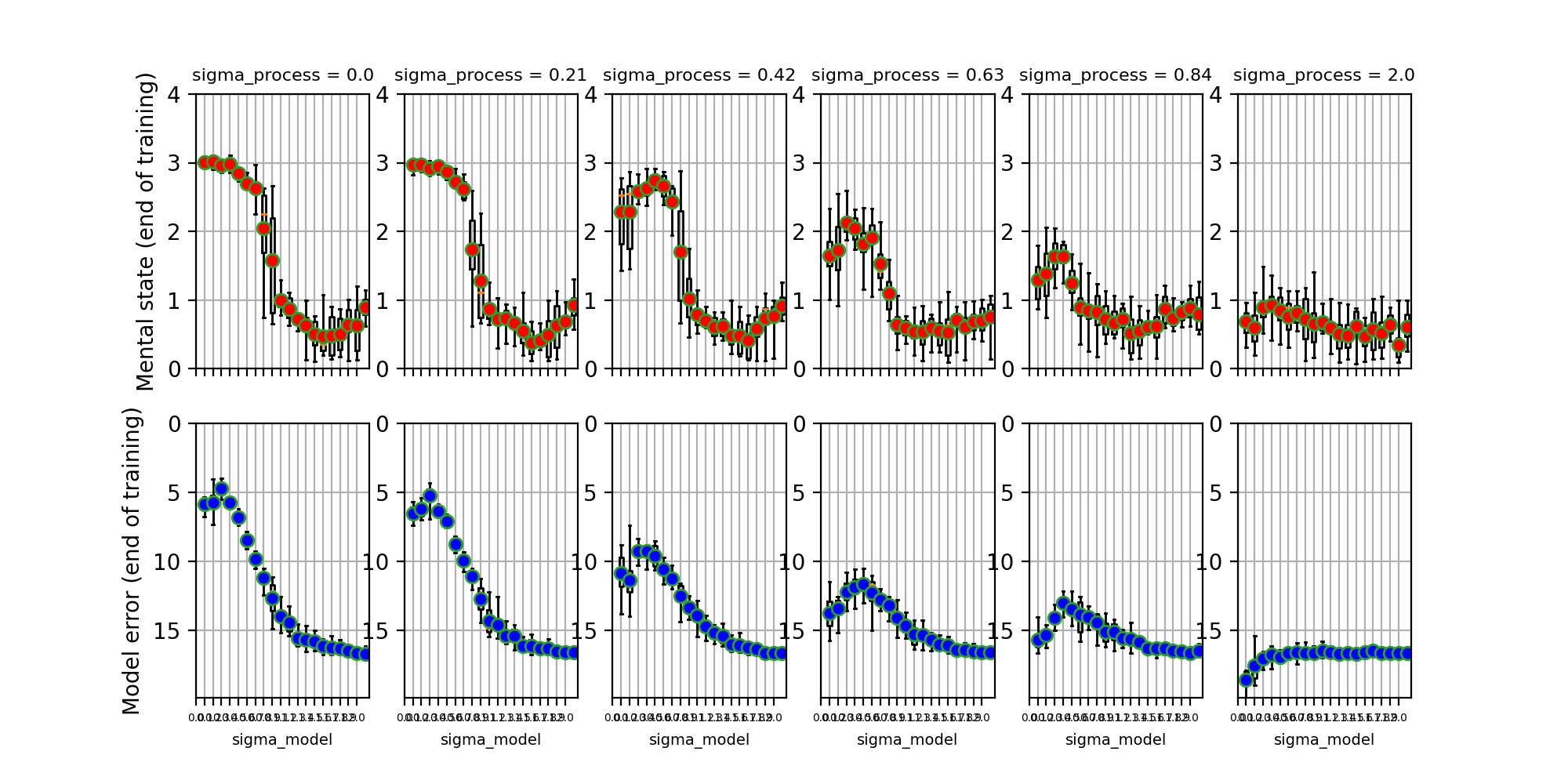}
        \caption{Optimal subject internal belief depending on feedback noise.}\label{fig:agent_double_var_nolearn_2}
    \end{subfigure}
    \caption{Simulated trials for 20x20 groups of 10 agents. Each group trained used a specific $(\sigma_{process},\sigma_{model}) \in [0.01,2.0]*[0.01,2.0]$ set of parameters.\ref{fig:agent_double_var_nolearn_1} shows the final average mental state (last 15 trials) and action model error (defined as the KL divergence between subject model $\mathbf{b}$ and the true mental transitions $\mathbf{B}$) for each of these groups. Figure \ref{fig:agent_double_var_nolearn_2} focuses on a few specific true feedback noise values and shows which subject prior model parameters lead to good regulation results. This simulation yields expected results (the training can only be useful if the feedback isn't too noisy, if the subjects trust the feedback somewhat, etc.), and less instinctive ones (subjects with some degree of caution $\sigma_{model}\simeq0.5$ learn better than naive ones whatever the actual level of feedback noise).} 
    \label{fig:agent_double_var_nolearn}
\end{figure*}

We provided artificial subjects with a diversity of feedback noises. In contrast with previous simulations, agents were cautious about the indication of this feedback. This meant that subject perception of their cognitive state was not a trivial endeavour. Figure \ref{fig:agent_cautious} shows the main difference with simulations of section \ref{sec:results_3.2.1} : the subjects beliefs about their mental state became much more uncertain, leading to less radical mental action learning. Indeed, the subjects uncertainty regarding the explicit feedback reflected on what they were able to learn from it : if they doubted it too much, it proved very difficult to develop an accurate model of interaction with the feedback. Despite that fact, a reasonable amount of caution towards the feedback did not lead to significant impairment regarding self-regulation for a perfect feedback. \ref{fig:agent_cautious_4}. 

In fact, cautious subjects ($\sigma_{model}=0.5$) performed better than  naive subjects ($\sigma_{model}=0.01$) when the feedback stood in the critical $\sigma_{process} \simeq 0.5$ range \ref{fig:compared_training_curve}. As one would expect, subjects who have a better understanding of the feedback tend to limit overfitting from noisy observations and build a better model of their mental actions. 

We generalized those findings to a wider set of true feedback noise / subject feedback noise expectations combinations to figure out which set of subject initial expectations would favor learning from even noisy feedback. Figure \ref{fig:agent_double_var_nolearn_1} shows performance metrics for 20 x 20 groups of 10 agents. Each group trained used a specific $(\sigma_{process},\sigma_{model})$ set of parameters. We show the final average mental state (last 15 trials), feedback model error and action model error for each of these groups. As expected, good learning only occurs when the feedback quality is good enough, and when the subjects actually believed that the feedback contained some useful information about their mental activity. More surprisingly, having a perfect model of the true feedback (i.e. knowing exactly the mapping from one's mental state to the observed feedback) did not always result in better mental action models. Finally, it appears that finding the optimal subject expectations with regard to the quality of the chosen biomarker is a balancing act. In the simple simulations we conducted, the optimal subject expectations parameter evolved dynamically with the quality of the feedback \ref{fig:agent_double_var_nolearn_2}. However, for small biomarker noises ($\sigma_{process} \in [0.3,1.0]$), $\sigma_{model} \simeq 0.5$ yielded the best training results, suggesting that when the feedback provided is reasonably noisy, cautious priors lead to better learning efficiency. The cautious subjects also performed well under perfect feedback, suggesting that under our approximations, a small degree of flexibility in the subject feedback perception led to overall better learning.

\subsection{Modelling changing subject confidence in the feedback}\label{sec:uncertainties_sim}

Previous simulations have shown how the noise affecting the biomarker affected the training of a subject depending on its expectations. However, they also assumed that those expectations were fixed and did not change accross the training. We believe this oversimplification glosses over one of the major mechanics of BCI training : the subject changing level of confidence about the reliability of the provided feedback accross the trials. The next simulations featured the same model as previous sections (\ref{sec:3.1} and \ref{sec:3.2}) with an important distinction : the subjects dynamically changed their observation mapping (relating cognitive state and observation), thus accounting for BCI training-specific dual perception-action uncertainty.

\begin{figure*}
    \centering
    \begin{subfigure}[b]{0.32\textwidth}
        \centering
        \includegraphics[width=\linewidth]{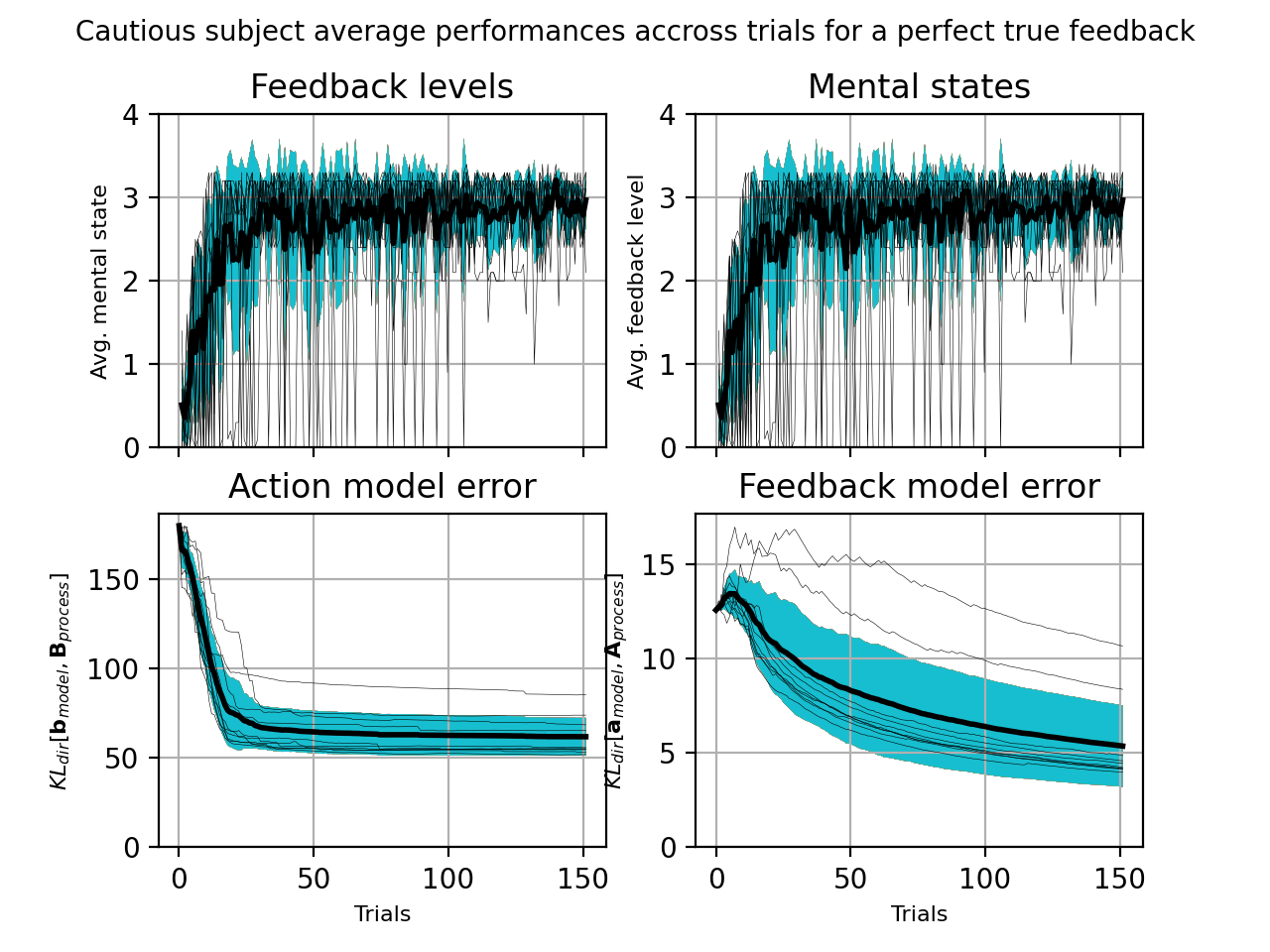}
        \caption{$\sigma_{process}=0.01,\sigma_{model}=0.5$}\label{fig:agent_double_var_learn1}
    \end{subfigure}
    \hfill
    \begin{subfigure}[b]{0.32\textwidth}  
        \centering 
        \includegraphics[width=\linewidth]{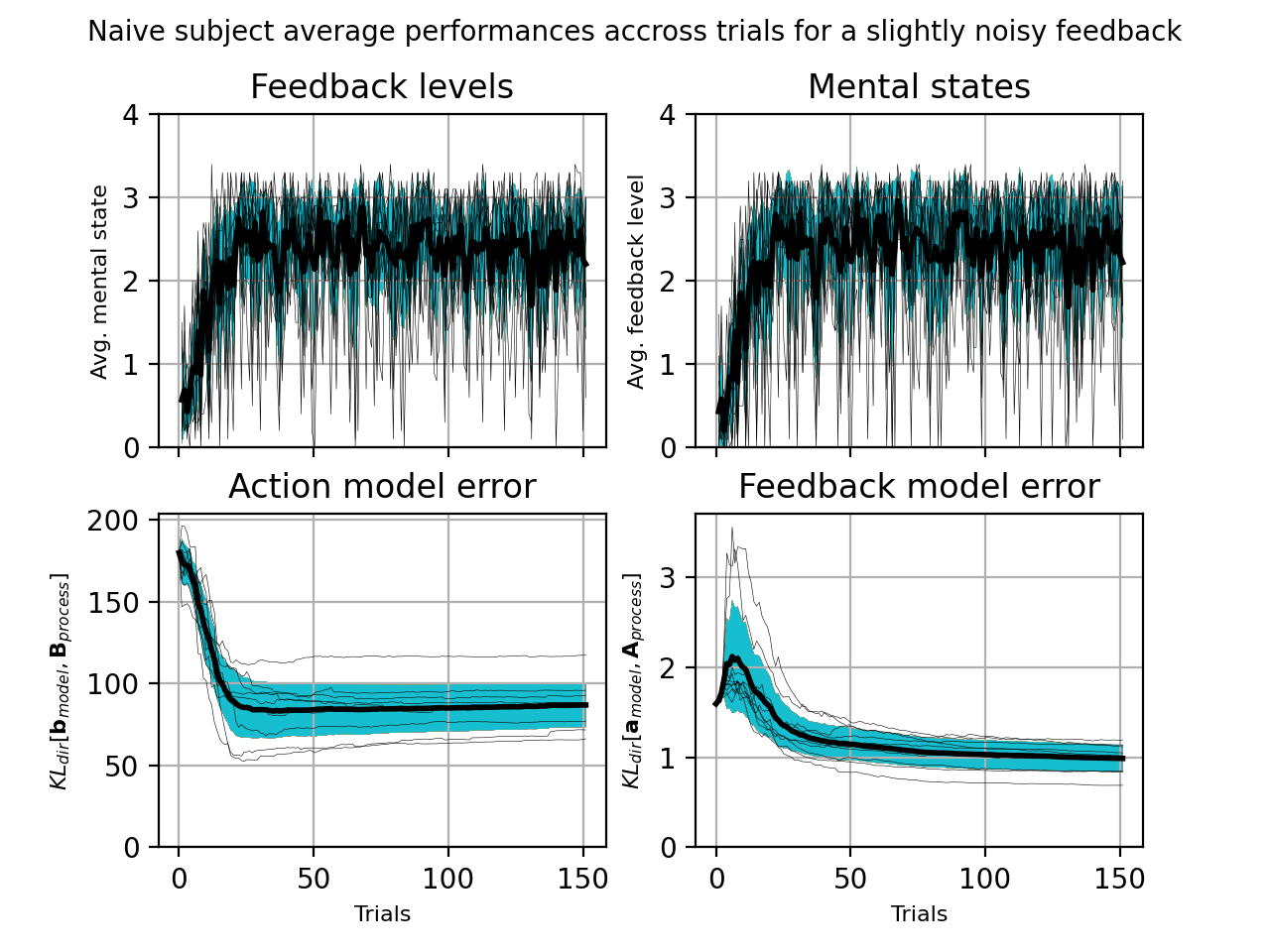}
        \caption{$\sigma_{process}=0.5,\sigma_{model}=0.01$}\label{fig:agent_double_var_learn2}
    \end{subfigure}
    \hfill
    \begin{subfigure}[b]{0.32\textwidth}   
        \centering 
        \includegraphics[width=\linewidth]{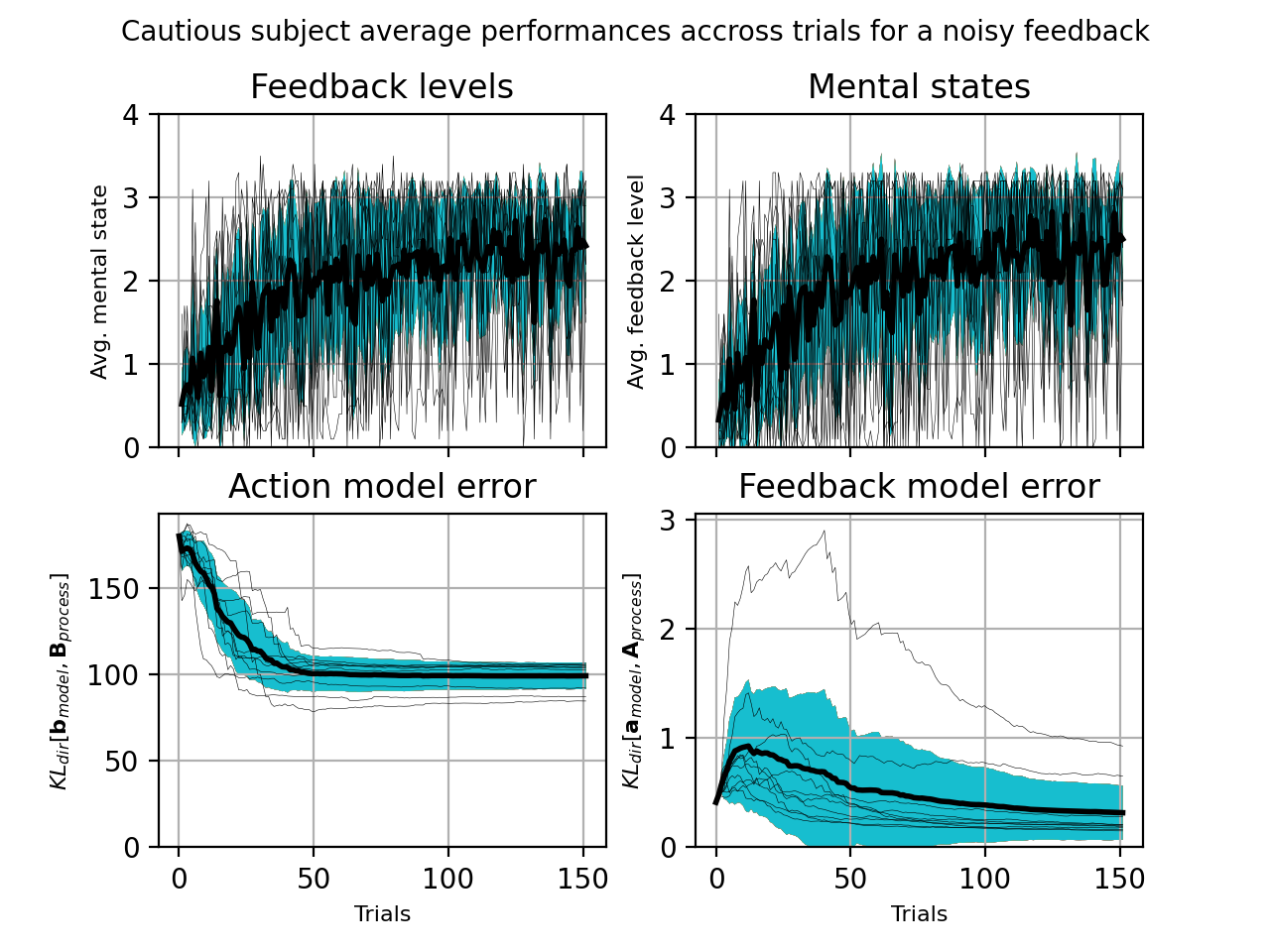}
        \caption{$\sigma_{process}=0.5,\sigma_{model}=0.5$}\label{fig:agent_double_var_learn3}
    \end{subfigure}
    \vskip\baselineskip
    \begin{subfigure}[b]{0.49\textwidth}   
        \centering 
        \includegraphics[width=\linewidth]{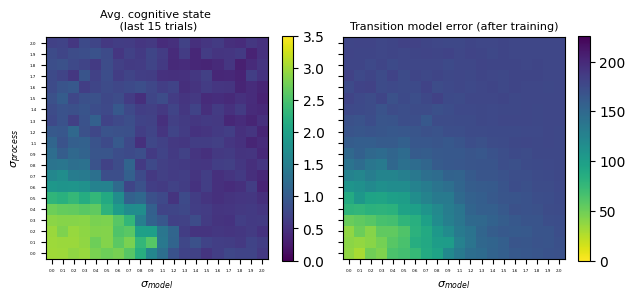}
        \caption{Agent performance map for a given expected feedback noise (x-axis) and true feedback noise (y-axis).}\label{fig:agent_double_var_learn4}
    \end{subfigure}
    \hfill
    \begin{subfigure}[b]{0.49\textwidth}   
        \centering 
        \includegraphics[width=\linewidth]{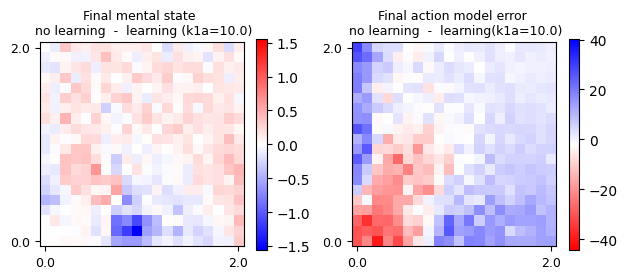}
        \caption{Final mental state and action model error differences between \ref{fig:agent_double_var_nolearn_1} and \ref{fig:agent_double_var_learn4}. Blue values means continuously updating your feedback belief led to better performances.}\label{fig:agent_double_var_learn4bis}
    \end{subfigure}
    \vskip\baselineskip
    \begin{subfigure}[b]{\textwidth}   
        \centering 
        \includegraphics[width=\linewidth]{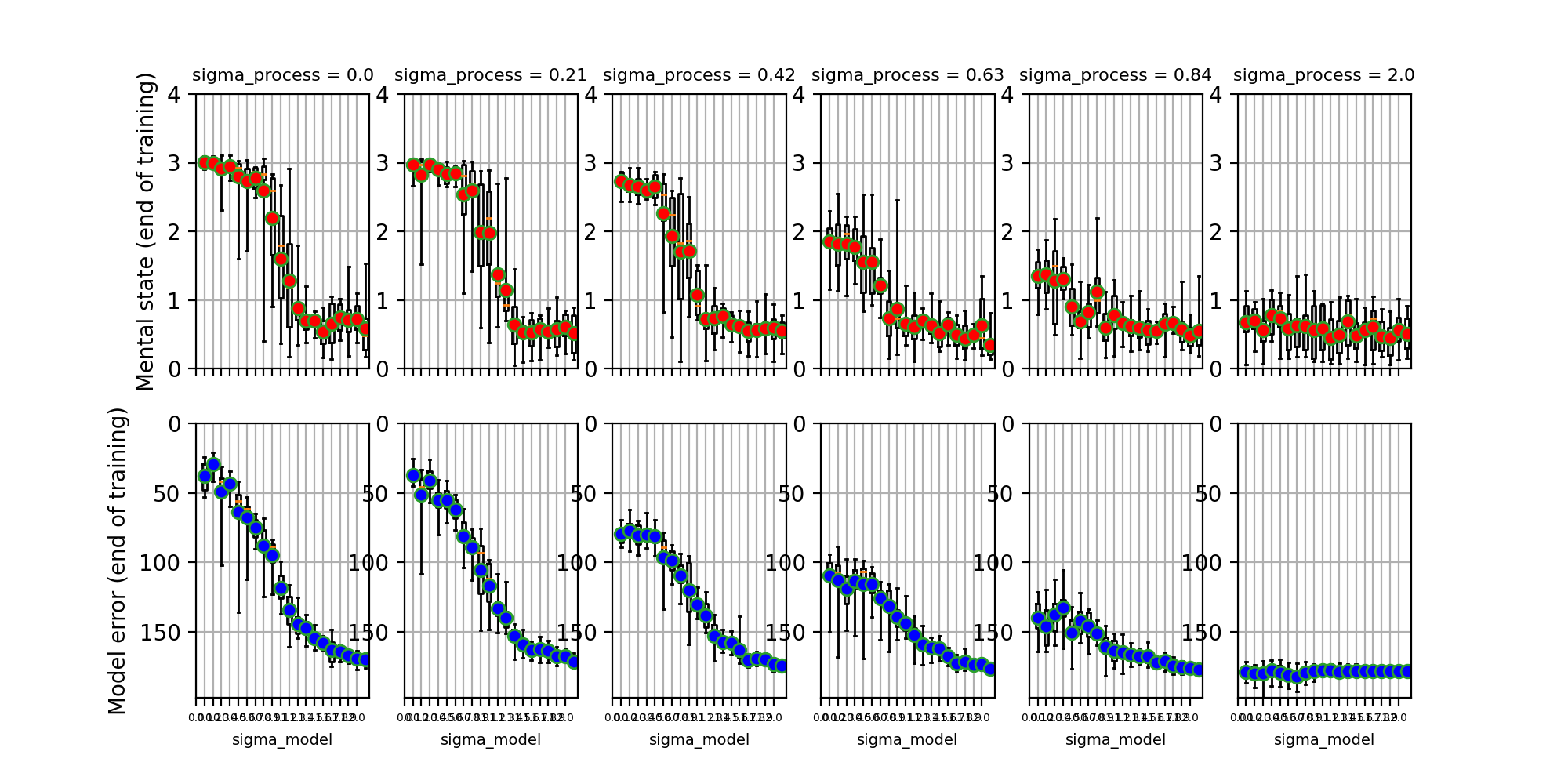}
        \caption{Optimal subject internal belief depending on feedback noise. Simultaneously learning action and feedback mapping leads to lower performances,higher within-group variance and the absence of the optimal subject caution prior noticeable in \ref{fig:agent_double_var_nolearn_2}.}\label{fig:agent_double_var_learn5}
    \end{subfigure}
    \caption{Simulated trials of agents which updated their model of the feedback and their action model simultaneously. The agents trusted their feedback priors ($k1_a = 10.0$) and dynamically updated their feedback model. We show the training curves for some typical cases (agent confidence / feedback quality mismatches) in \ref{fig:agent_double_var_learn1},\ref{fig:agent_double_var_learn2},\ref{fig:agent_double_var_learn3}. We generalize these results to  $(\sigma_{process},\sigma_{model}) \in [0.01,2.0]*[0.01,2.0]$ in \ref{fig:agent_double_var_learn4}. Agents updating their feedback model led to overall lower performances compared to \ref{fig:agent_double_var_nolearn}. Our model suggests that for lower values of $k1_a$, it was easier for subjects to doubt the reliability of the feedback signal rather than learning complex mental actions.} 
    \label{fig:agent_double_var_learn}
\end{figure*}

We show the results of these simulations in figure \ref{fig:agent_double_var_learn}. First, we simulated very typical training setups, namely a cautious subject provided with a perfect feedback \ref{fig:agent_double_var_learn1}, a naive subject provided with a noisy feedback \ref{fig:agent_double_var_learn2} and a cautious subject provided with a noisy feedback \ref{fig:agent_double_var_learn3}. In an effort to reduce their Free Energy, the agents updated their models of the feedback (bottom right plots). Because the agents learnt action and perception models simultaneously, those learning trajectories affected one another dynamically. In general, this manifested as a deterioration of the feedback model while the action model improved. Once the mental action topography was sufficiently explored, the models of the agents converged towards local free-energy minima.Figure \ref{fig:agent_double_var_learn4} generalizes these remarks to a wider range of true feedback / subject expectation parameters. 

Finally, we wanted to know if questioning the quality of the feedback provided negatively impacted  BCI training subjects. We compared the performances of these feedback-learning agents to the results showed in \ref{fig:agent_double_var_nolearn_1} (fixed feedback mappings). Our simulations (\ref{fig:agent_double_var_learn4bis}) showed that the impact actually varied based on the subject's initial model : when the subject's model initially closely aligned with the true feedback process, continuously learning the feedback tended to hurt the performances of the training? In turn, for subjects with feedback mappings significantly divergent from the ground truth (such as naive or overly cautious subjects), learning the feedback resulted in better action mappings, displaying some kind of 'protection mechanism' preventing subjects from learning damaging mental actions. 

\subsection{Simulating training with Interoceptive Learning}\label{sec:4.4}

\begin{figure*}
    \centering
    \begin{subfigure}[b]{0.49\textwidth}
        \centering
        \includegraphics[width=\linewidth]{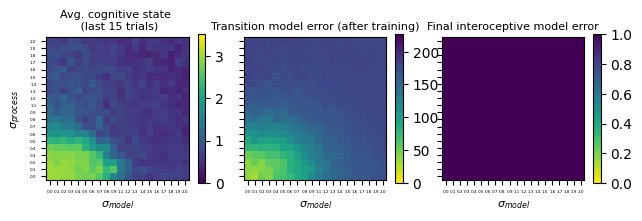}
        \caption{No interoceptive observation modality.}\label{fig:agent_io_1}
    \end{subfigure}
    \hfill
    \begin{subfigure}[b]{0.49\textwidth}  
        \centering 
        \includegraphics[width=\linewidth]{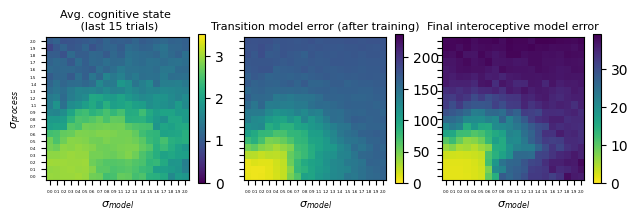}
        \caption{$\sigma_{intero}\simeq0.0$}\label{fig:agent_io_2}
    \end{subfigure}
    \vskip\baselineskip
    \begin{subfigure}[b]{0.49\textwidth}   
        \centering 
        \includegraphics[width=\linewidth]{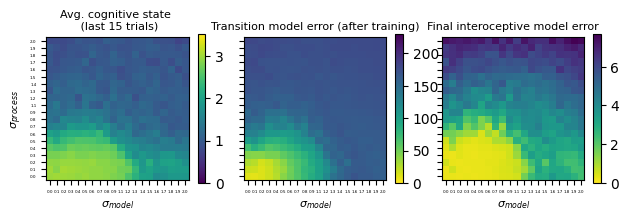}
        \caption{$\sigma_{intero}\simeq0.9$}\label{fig:agent_io_3}
    \end{subfigure}
    \hfill
    \begin{subfigure}[b]{0.49\textwidth}   
        \centering 
        \includegraphics[width=\linewidth]{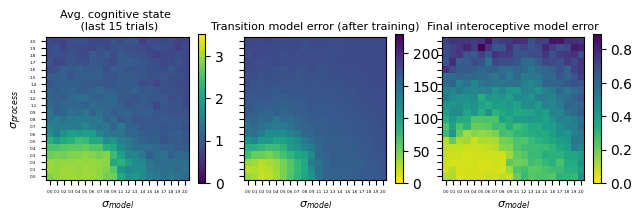}
        \caption{$\sigma_{intero}\simeq2.0$}\label{fig:agent_io_4}
    \end{subfigure}
    \caption{Simulated training results for four groups of agents. Agents of group 1 \ref{fig:agent_io_1} had no interoception (just like in figure \ref{fig:agent_double_var_learn4}). Agents of groups 2,3,4 (\ref{fig:agent_io_2},\ref{fig:agent_io_3},\ref{fig:agent_io_4}) were all equipped with an interoceptive observation modality characterized by a specific noise ($\sigma_{intero}\simeq0.0$,$\sigma_{intero}\simeq0.9$ and $\sigma_{intero}\simeq2.0$ respectively). Each group was subdivided in 20 x 20 subgroups of 10 agents, each using a specific $(\sigma_{process}, \sigma_{model})$ set of parameters. The figures show the final average mental state (last 15 trials), learnt action model error and learnt interoceptive model error for each subgroup of agents.} 
    \label{fig:agent_io}
\end{figure*}

We ran simulations where agents were provided with an additional (interoceptive) observation modality. Simulated subjects internal signals were either informative (low $\sigma_{model}^{intero}$) or uninformative (high $\sigma_{process}^{intero}$). Figure \ref{fig:agent_io} shows the performance maps of agents with various degrees of internal observation abilities. Overall, agents equipped with an internal observation modality outperformed agents relying on the external feedback only. Particularly noteworthy is the superior performance of subjects with precise internal signals, as depicted in Figure \ref{fig:agent_io_2}, where they demonstrated the ability to decipher high-noise feedback. This suggests that the substantial decline in training efficiency, as evident in the alarming drop shown in Figure \ref{fig:biomarker_noise_plot}, might be alleviated by leveraging this alternative observation pathway. Importantly, the efficacy of this supplementary observation route is contingent upon the quality of the interoceptive pathway.

\section{Discussion}\label{sec:discuss}

The aim of this paper was two-fold : first, showing how the cognitive system of the subject could help describe the complex regulation mechanisms at work during Neurofeedback Training. Second, demonstrating how the Active Inference framework could provide a pertinent set of tools to computationally model training and contribute to BCI research by testing computationally various hypotheses and the effect of possible instructions. 

\subsection{Cognitive regulation at the heart of Neurofeedback learning}

In usual neurofeedback mechanism accounts, brain activation has long been considered as the independent variable, with cognition and behaviour being dependent variables \cite{sitaram_closed-loop_2017}. We suggest an alternative framing approach based on subject cognitive control of the feedback. In such a formulation, the subject guides the dynamics of the training through perception, mental actions and learning. In this formulation, brain activity is partially predicted by cognitive dynamics, in turn dependent on high level perception, planning and learning. Note that these processes need not be \textit{conscious} for the subject. In over words, our model thus does not extend explicitly to the physiological dimension of subjects' brains, but focuses on the cognitive states they are coupled with. Of course, the assumed link between the modeled cognitive states and the actual measured biomarkers remains a critical (and often weak) point in practice, reinforcing the importance of biomarker selection and proper reporting \cite{ros_consensus_2020}. \\

\textbf{Ability to generalize :} Training success or failure is a difficult concept to formalize clearly, mainly because the appreciation of a successful training varies heavily depending on the paradigm. Moreover, the ability of subjects to transfer the knowledge gathered from training into situations outside of the training environment is crucial for the success of BCI / neurofeedback training, especially in therapeutic approaches. On the one hand, a general argument can be made that BCI training is in essence a cognitive exercise aiming to provide the subjects with a way to learn adequate mental strategies. Following this argument, an untrained subject lacks a cognitive model accurate enough to navigate its (mental) environment easily. This may lead to observable behavioural afflictions, or the inability to control an external device, etc. Training would allow subjects to build an accurate model of perception and action, making them able to control their cognition more willingly, or build reliable internal observers (metacognition). Thus, training successfully would be synonym to building a good internal model. Following our Active Inference formulation, this would mean that a successful training would minimize the difference between the true environment dynamics $\mathbf{A}$, $\mathbf{B}$ and the modeled dynamics $\mathbf{a}, \mathbf{b}$ in the targeted cognitive dimension.

Another prevalent hypothesis postulates the learning of an alternative, internal observation modality : other modeling approaches such as \cite{davelaar_mechanisms_2018} have already theorized the emergence of an interoceptive observation pathway during training and discussed their ability to reinforce an ongoing training and then be used as the only reinforcer after training. Our model show this reinforcing effect during training (see simulation \ref{sec:4.4}), and underlines the importance of this "gut-feeling" as it would render the training useless to the subject when cut from the external feedback. In the initial simulations, we greatly simplified the nature of subject experience during training. Indeed, we assumed that the only way for the subject to infer their cognitive activity was to use an external (BCI-equipped) observation pipeline. Although this representation may suffice when the end-goal of training is learning how to use an external device, it falls short when tackling neurofeedback generalization problems : when deprived of the external feedback, how would the subject perform self-regulation ? We believe that empirical problem is solved through learning an \textit{internal} observation modality. We modeled this interoceptive pathway as an independent observation modality, similar to the external feedback, but other representations, such as hierarchical models \cite{hesp_deeply_2021,sandved-smith_towards_2020}, may provide promising alternatives.\\

\textbf{Tool learning \& illiteracy :} Although the task the synthetic agents were given would have been trivial if the subjects knew all the dynamics of the environment (\textit{What is the effect of my actions on my mental states ?} and \textit{How does a given mental state translate to a feedback level ?}), it proved much harder when they were facing uncertainty from one (or both) modalities. Learning to interact with a new tool requires the user to build a comprehensive model of observations and actions. During BCI interaction this is particularly difficult, because subjects beliefs about observation and action modalities lack the prior evidence needed to dissipate this uncertainty.  To learn effectively, the subject needs to be quite confident in part of its model, in order to progressively design better and better interaction models. Our simulations have shown that high amount of uncertainties in the subject model led to poor training performances, even in the advent of a precise feedback.  This is true in every environment (indeed, in the Active Inference literature, most approaches focus on learning either observation matrices given known action matrices \cite{friston_active_2016,smith_step-by-step_2021,Tschantz2020} or the contrary.) In our approach however, because neurofeedback and other BCI-based applications rely on a much more uncertain set of dynamics, the model we propose will feature various levels of uncertainty in both agent perception and action beliefs. The complexity of the inference problem and its dependence to subject initial priors is a first explanation for high inter-subject heterogeneity.\\
Finally, our simulations suggest that in some cases, it may be easier for the subject to explain their own poor training performances by questioning the reliability of the feedback provided (regardless of its actual intrinsic quality), rather than their own failure to figure out the adequate transition rules. Simulated agents who continuously updated their confidence in the feedback performed worse than their more naive counterparts when the feedback reliability was high and matched their initial beliefs, but better when the feedback was of poorer quality. It remains to be seen wether such fine training effects may be captured using the data from usual BCIT tasks.

\subsection{Modeling BCI training}
The Active Inference framework proves particularly suitable for modelling Brain-Computer Interface (BCI) training for several reasons:
\begin{itemize}
    \item It accommodates subject-specific internal parameters and learning environment criteria effectively.
    \item	It encompasses both action, perception and learning, essential components of BCI training dynamics.
    \item	Its versatility enables the modelling of various types of training loops and subject models within a unified framework. The generic nature of the model described in the paper means that the framework accounts for the great diversity in BCI training protocols (biomarkers, feedback design, trained cognitive dimension).
    \item	Possible extensions of the model offers scalability and hierarchical structuring, accommodating the complexity of BCI training scenarios. \cite{Friston2017hierarchical,Pezzulo2018hierarchical}
    \item	The framework reflects various timescales, encompassing quick processes such as state inference and decision making, and slower ones like learning.
\end{itemize}

By leveraging Active Inference, it becomes possible to conceptualize subject behaviour during BCI training as akin to navigating through an uncertain environment. Crucially, this formalism incorporates numerous initial and hyper-parameters, empowering modelers to replicate the reality of BCI training. Manipulating these parameters enables the exploration of the potential limitations of specific training protocols and their effects on agent behaviour. 
Of particular significance are the biomarker reliability (including bias and noise) underlying the feedback and the priors within the subject model (representing the subject's initial beliefs about the system at the onset of training). Our basic simulations demonstrate that a perfectly designed feedback mechanism alone does not ensure successful training. Consequently, we can establish links between fundamental elements of the BCI training protocol \cite{ros_consensus_2020} and information theory-based variables to investigate learning efficacy.
The broad applicability of the Active Inference framework presents both advantages and challenges. On one hand, its generic nature allows for the modeling of a diverse array of situations using the same set of tools. On the other hand, this versatility necessitates a wide range of parameters, making the model difficult to test and ultimately challenging to fit. To restrict the space of simulated parameters, we imposed constraints on the initial values of these parameters. Specifically, we:
\begin{itemize}
    \item Assumed the true topography of the mental states of the subject and their representation in the subject model. We also chose to limit state transitions to neighboring mental states in order to describe a continuous process. Finally, we used spontaneous state-transitions towards lower states to mirror a mental resting state.
    \item Modeled the Feedback / biomarker as noisy, but unbiaised, in effect refraining from delving into the expansive model space of specific biomarker biaises, as their effects may prove trivial and difficult to thoroughly examine. We also described the subject initial feedback priors noisy but unbiaised for the same reason. 
\end{itemize}

\textbf{Explaining BCI training variability : }In our study, one of the primary objectives was to elucidate the inter-subject differences observed in certain studies, where some participants succeeded in regulating their brain activity while others did not, leading to what has been termed "BCI illiteracy". Within our formalism, these individual differences can be attributed to several factors. Firstly, within a single group of agents with similar internal parameters, the stochastic nature of the biomarker noise, and the decision-making process of our agents may introduce randomness, particularly when multiple actions exhibit similar perceived value. As a result, simulating the behaviour of a single agent with specific characteristics did not suffice, necessitating the examination of mean performance indicators across multiple agents with similar initial properties to draw meaningful conclusions. 

Our proposed formalization underscores the structural sources of uncertainty inherent in generic BCI training paradigms, emphasizing the complexity of the training task due to high uncertainty in both perception and action domains. Typically, when faced with uncertainty regarding a set of dynamics, strong priors in other dimensions can facilitate the interpolation and construction of a coherent model of the uncertain dimension. However, in many BCI applications, including neurofeedback, agents often start with uncertain priors in both perception and action, making it challenging to establish the relationship between feedback and cognitive states. This dual uncertainty offers a plausible mechanistic explanation for the variability observed in neurofeedback efficacy studies.

Furthermore, discrepancies in initial parameters can contribute to group effects in the outcome of the training process, with variations stemming from factors such as motivation, habits, and prior knowledge. Our model allows researchers to shed light on the impact of these group differences by considering alternative preference matrices, action habits, and prior models.

Nevertheless, our simulations demonstrate that under appropriate conditions, the challenges posed by this uncertainty may be addressed, and successful model learning and system navigation may be achieved even with imperfect priors. Simulated trainings with the additionnal interoceptive observer proved much more succesful. Agents equipped with this additionnal observatory pathway required less demanding priors qualities to achieve satisfactory outcomes.

\subsection{What lessons for BCI experimenters ? }

Our basic simulations have underlined the general dynamics of some parameters of neurofeedback training. First, we showed the importance of the reliability of the biomarker used : the threshold noise above which the subjects became unable to self-regulate was especially low with a brutal fall-off. This indicated that neurofeedback training paradigms based on nonspecific biomarkers may face tough odds. In general, the ability of the subjects to make sense of the feedback was directly related to how discriminant it was, i.e. how different two distinct mental states manifested themselves in the feedback. 

The priors of the subjects regarding the feedback and their mental strategy played a particularly important role in our simulations. Although these priors may depend on subject/population-specific parameters, they are heavily influenced by experimenter instructions. In general, a normalization and clarification of the instructions given to the subjects has long been called for \cite{ros_consensus_2020} and constitutes an important first step to (1.) simplify the patient task, (2.) allow for better reproductibility, (3.) better explain the outcome of training. A more well-defined set of training parameters may also be used to better understand the causes of training failure and hopefully, solve them. More precisely though, basic simulations showed that :\\
\begin{itemize}
    \item \textbf{Feedback :} In general, agents benefited from a limited amount of prior caution regarding the feedback, i.e. not being too naive about its accuracy. Doubting the feedback (\ref{sec:uncertainties_sim}) acted as a protection mechanism in very poor conditions (i.e. when subject priors were significantly off or when feedback noise was significantly high) to avoid learning flawed correlations. However, it also hurt the performances of subjects with as initially adequate feedback perception, suggesting that experimenters should honestly communicate about the reliability of their NFT paradigms, though the nature of this communications remains vague. 
    \item \textbf{Mental strategies:} In our simulations, we deliberately restricted the number of available mental actions to facilitate learning within a feasible timeframe. Prompting agents to explore a specific subset of their mental actions (through low initial action mapping confidence) was crucial for initiating the necessary process of exploration and learning. This suggests that pushing subjects to try out specific mental strategies at the start of the training is particularly important, and may prevent training if overlooked. This is a debated issue in the litterature, with some arguing that providing a clear set of strategies and showing examples may promote the learning of the BCI skill \cite{Lotte2013}, while others warn against vague instructions that may be incongruent to the desired outcome \cite{Eddy2016}.
    \item \textbf{Interoception :} We described subjects ability to perceive their own mental states without the feedback as an independent internal observation modality . More than just an reinforcing mechanism during training, we argue it is actually a necessary pathway to allow subjects to use the knowledge acquired during neurofeedback training once deprived of the external feedback. To favor learning this \textit{gut intuition}, the experimenter instructions should drive the subjects towards attending internal signals. More generally, this interoceptive observation pathway has been cited as a prime candidate for post-training effect retention, by allowing homeostasis \cite{davelaar_mechanisms_2018}.
\end{itemize}

\subsection{Modeling limits}
Although our models of BCI interaction are, necessarily, greatly simplified, a few limitations are worthy of a mention, both within and beyond the perimeter of our approach :
\begin{itemize}
    \item Physiological activity of the subject : Brain Activity regulation has been at the center of Neurofeedback training, with some studies assessing the success of their training based on the degree physiological regulation (while sometimes forgetting to check its effect on subject behaviour). Some pre-existing modeling approaches have explicitly described its evolution through regulation. We chose to render the dimension implicit by providing a direct function mapping cognition and feedback. Simulating the NF subject physiology is a necessary next step for our model to be deployed and used in actual BCI practice.
    \item Subject observable behaviour during and after training : Although cognitive actions directly related to the feedback are modelled, observable behaviour is  not simulated. Of course, this is a major limit of our approach as it prevents us from comparing the results of regulation and behaviour (the usual end goal of neurofeedback training).
    \item The temporal dynamics of the feedback training loop  : Previous neurofeedback modeling approaches have tackled the influence of temporal factors such as delays or continuous feedback on training \cite{oblak_2017} . However, due to the sequential nature of our framework, it may not be fit to explore the effect of feedback temporality on human perception on its own.
    \item Subject behaviour outside of the training environment. This thematic is particularly important when talking about therapeutic neurofeedback. The effect of NFT on the subjects outside of the training sessions proper is not simulated directly : we can only make simplifying assumptions regarding the relationship between agents models after training (learning, habits, etc.) and their performances in other situations. This would most likely be linked to concepts such as transfer learning that we don't consider here.
\end{itemize}
Those limits prevent us from accounting for some dimensions of the training. Additionally, some hypotheses have been made regarding the properties of the subject generative model and cognitive states. Those hypotheses are summed-up in section \ref{sec:active_inference} and tend to make (reasonable) assumptions on perception and action selection rules during neurofeedback.

\subsection{Next leads}

The work presented here paves the way for three main developmental axes. 

First, Active Inference models are good candidates to be used in (Bayesian) parameter estimation methods \cite{Jaakkola2000}. This would allow researchers to use experimental training data to figure out the most likely subject / training pipeline characteristics to explain their results. Beyond better result interpretation, proven training models may be used as the basis for training outcome predictors or adaptive training protocols. 

Second, equipping the model with a neurophysiological component in order to explicitly feature the biomarker used during training would allow us to \textit{close the loop} and propose a more complete approximation of  the coupling between cognition and physiology in the brain. This may eventually allow us to confront alternative models to experimental results  in order to more closely fit subject internal models as they interact with the BCI. Although tricky, this could be achieved by using the predicted synthetic physiological responses generated during belief update \cite{friston_active_2016}. Alternatively, adapting compatible methods used in signal classification such as deep autoregressive Hierarchical Markov Model \cite{Wang2018} may prove interesting.

Third, possible extensions of the current (cognitive) model to specific cognitive dynamics such as attention may prove instrumental to study the influence of (meta-)cognitive dimensions such as attention. Active Inference based approaches such as \cite{hesp_deeply_2021,sandved-smith_towards_2020} have already tackled the question of attention and valence in the Active Inference framework, and models of self-regulation using this set of models may yield more specific training curves for neurofeedback training paradigms targetting ADHD.

\section{Concluding remarks}

This paper seeks to contribute to the neurofeedback efficacy debate by proposing a new modeling angle, namely the cognitive system of the subject. Importantly, our study makes the brain activity of the subject implicit. We propose a generic family of models to characterize most brain-computer interface training paradigms as complex inference problems that subjects must solve through perception, action and learning despite uncertain priors and feedback. This formulation allows researchers to explicitly model the effect of key variables such as the prior confidence of the subjects, the influence of experimenter instructions, the quality of the feedback or the motivation of the subjects, where previous physiology-based modeling approaches struggled.

We utilize the Active Inference framework to simulate the behavior of biologically plausible artificial agents tasked with solving our proposed problems. This endeavor enables experimentation on thousands of synthetic subjects, allowing us to estimate the outcomes of basic experimental protocols without relying on time-consuming and often underpowered studies. 

The predictions derived from these minimal models were used to provide a first quantitative account of the influence of biomarker noise, subject expectations and learning mechanisms in neurofeedback training. Finally, this mechanistic account of subject experience suggests that developing a precise interoceptive signal not only makes the training easier but may also be essential in order to generalize the effect of neurofeedback training to everyday life environments.

\section{Acknowledgements}

The authors thank L. Da Costa for the insightful discussions.

\section{Competing interests}

The authors declare no competing interests.

\section{Ressources}

Computations of the Active Inference agents were made on a custom Python implementation of the SPM12 toolbox Matlab \textbf{MDP\_VB\_X.m} and \textbf{MDP\_VB\_XX.m} scripts (available at \url{https://www.fil.ion.ucl.ac.uk/spm/}) with minimal changes. The custom package used is publicly available at url: \url{https://github.com/Erresthor/ActivPynference_Public}.

\bibliographystyle{alpha}
\bibliography{my_bib,sample}

\end{document}